\documentclass[useAMS,usenatbib]{mn2e}

\usepackage[letterpaper,margin=0.635in]{geometry}
\usepackage[ansinew]{inputenc}
\usepackage{amsfonts}
\usepackage{amsmath}
\usepackage{enumerate}
\usepackage{enumitem}
\usepackage{graphicx}
\usepackage{epsfig}
\usepackage{color}
\usepackage[normalem]{ulem}
\usepackage{comment}
\usepackage{amssymb}
\usepackage{pdflscape}
\usepackage{float}
\usepackage{booktabs}

\usepackage[pagewise]{lineno}

\DeclareMathAlphabet{\mathscr}{OT1}{pzc}%
                                 {m}{it}

\def\aap{{ A\&A}}

\def\aj{{AJ}}

\def\apj{{ApJ}}
\def\apjl{{ApJL}}

\def\mnras{{MNRAS}}

\def\jcap{JCAP}

\def\redmagic{redMaGiC}

\newcommand{\prd}{Phys. Rev. D}
\newcommand{\physrep}{PhR}
\newcommand{\procspie}{SPIE}
\newcommand{\ssr}{Space Science Reviews}

\newcommand{\gamt}{\gamma_t(\theta)}

\newcommand{\angstrom}{\mbox{\normalfont\AA}}

\newcommand{\wtheta}{w(\theta)}
\newcommand{\mr}{\mathrm}

\def\ave#1{\left\langle #1 \right\rangle}

\newcommand{\zl}{z_{\rm L}}
\newcommand{\zs}{z_{\rm s}}

\newcommand{\be}{\begin{equation}}
\newcommand{\ee}{\end{equation}}
\newcommand{\bes}{\begin{equation*}}
\newcommand{\ees}{\end{equation*}}
\newcommand{\bea}{\begin{eqnarray}}
\newcommand{\eea}{\end{eqnarray}}
\newcommand{\beas}{\begin{eqnarray*}}
\newcommand{\eeas}{\end{eqnarray*}}

\newcommand{\imshape}{{\textsc{im3shape}}}
\newcommand{\ngmix}{\texttt{ngmix}}

\title[Cosmology from the Dark Energy Survey]{Cosmology from large scale galaxy clustering and galaxy-galaxy lensing with Dark Energy Survey Science Verification data}
\author[Kwan, S{\'a}nchez et al.]
{\parbox{\textwidth}{
J.~Kwan$^1$\footnotemark,
C.~S\'{a}nchez$^2$\footnotemark, 
J.~Clampitt$^1$, 
J.~Blazek$^3$, 
M.~Crocce$^{4}$, 
B.~Jain$^1$, 
J.~Zuntz$^5$, 
A.~Amara$^6$, 
M.~R.~Becker$^{7,8}$, 
G.~M.~Bernstein$^1$,
C.~Bonnett$^2$, 
J.~DeRose$^{7,8}$, 
S. Dodelson$^{9,10,11}$, 
T.~F.~Eifler$^{12}$, 
E.~Gaztanaga$^4$, 
T.~Giannantonio$^{13,14}$, 
D.~Gruen$^{8,15}$, 
W.~G.~Hartley$^{16}$, 
T.~Kacprzak$^{16}$, 
D.~Kirk$^{17}$, 
E.~Krause$^8$, 
N.~MacCrann$^{5}$, 
R.~Miquel$^{2,18}$, 
Y.~Park$^{19}$, 
A.~J.~Ross$^3$, 
E.~Rozo$^{19}$, 
E.~S.~Rykoff$^{8,15}$, 
E.~Sheldon$^{20}$, 
M.~A.~Troxel$^5$, 
R.~H.~Wechsler$^{7, 8, 15}$,  %% end of principal authors
T. M. C.~Abbott$^{21}$,
F.~B.~Abdalla$^{17,22}$,
S.~Allam$^{9}$,
A.~Benoit-L{\'e}vy$^{23,17, 24}$,
D.~Brooks$^{17}$,
D.~L.~Burke$^{8,15}$,
A. Carnero Rosell$^{25, 26}$,
M.~Carrasco~Kind$^{27, 28}$,
C.~E.~Cunha$^8$,
C.~B.~D'Andrea$^{29,30}$,
L.~N.~da Costa$^{25,26}$,
S.~Desai$^{31,32}$,
H.~T.~Diehl$^{9}$,
J.~P.~Dietrich$^{31,32}$,
P.~Doel$^{17}$,
A.~E.~Evrard$^{33,34}$,
E.~Fernandez$^2$,
D.~A.~Finley$^{9}$,
B.~Flaugher$^{9}$,
P.~Fosalba$^2$,
J.~Frieman$^{9,10}$,
D.~W.~Gerdes$^{34}$,
R.~A.~Gruendl$^{27,28}$,
G.~Gutierrez$^{9}$,
K.~Honscheid$^{3,35}$,
D.~J.~James$^{21}$,
M.~Jarvis$^1$,
K.~Kuehn$^{36}$,
O.~Lahav$^{17}$,
M.~Lima$^{37,25}$,
M.~A.~G.~Maia$^{25,26}$,
J.~L.~Marshall$^{38}$,
P.~Martini$^{3,39}$,
P.~Melchior$^{40}$,
J.~J.~Mohr$^{31,32,41}$,
R.~C.~Nichol$^{29}$,
B.~Nord$^{9}$,
A.~A.~Plazas$^{12}$,
K.~Reil$^{15}$,
A.~K.~Romer$^{42}$,
A.~Roodman$^{8,15}$,
E.~Sanchez$^{43}$,
V.~Scarpine$^{9}$,
I.~Sevilla-Noarbe$^{43}$,
R.~C.~Smith$^{21}$,
M.~Soares-Santos$^{9}$,
F.~Sobreira$^{44,25}$,
E.~Suchyta$^1$,
M.~E.~C.~Swanson$^{28}$,
G.~Tarle$^{34}$,
D.~Thomas$^{29}$,
V.~Vikram$^{45}$,
A.~R.~Walker$^{21}$}
  \vspace{0.4cm}\\~\\
\parbox{\textwidth}{\centering \textsc{\Large(The DES Collaboration)} \\ \centering \textit{Author affiliations are listed at the end of this paper}\\ }}
\begin{document}

%\linenumbers
%\renewcommand\linenumberfont{\normalfont\tiny}

%\date{Accepted 1988 December 15. Received 1988 December 14; in original form 1988 October 11}

\pagerange{\pageref{firstpage}--\pageref{lastpage}} \pubyear{0000}

\maketitle

\label{firstpage}

\begin{abstract}
We present cosmological constraints from the Dark Energy Survey (DES)
using a combined analysis of angular clustering of red galaxies and their 
cross-correlation with weak gravitational lensing of background galaxies. 
We use a 139 square degree contiguous patch of DES data from
the Science Verification (SV) period of observations. %, accounting for
%less than 3\% of the full DES survey coverage. 
Using 
%a combination of
%angular galaxy clustering and galaxy-galaxy lensing of red galaxies on
large scale measurements, we constrain the matter density of the Universe
as $\Omega_m = 0.31 \pm 0.09$ and the clustering amplitude of the matter
power spectrum as $\sigma_8 = 0.74 \pm 0.13$ after marginalizing over seven
nuisance parameters and three additional cosmological parameters. 
This translates into S$_8 \equiv \sigma_8(\Omega_m/0.3)^{0.16} = 0.74 \pm 0.12$ for our
fiducial lens redshift bin at 0.35 $<z<$ 0.5, while S$_8 = 0.78 \pm 0.09$ 
using two bins over the range 0.2 $<z<$ 0.5. 
We study the robustness of the results under changes in the data vectors,
modelling and systematics treatment, including photometric redshift and shear
  calibration uncertainties, and find consistency in the
derived cosmological parameters. We show that our results are consistent
with previous cosmological analyses from DES and other data sets
and conclude with a joint analysis of DES angular clustering and galaxy-galaxy
lensing with Planck CMB data, Baryon Accoustic Oscillations and 
Supernova type Ia measurements. 
  %them to be
  %in agreement... to improve current constraints by... }
  %Scott:Emphaize combining after checking consistency 
%  About X\% of the estimated error
%  bars comes from marginalizing over systematic parameters. 

\end{abstract}

}
%\twocolumn[\head]

%%%%%%%%%%%%%%%
\section{Introduction}
\label{sec:intro}
%%%%%%%%%%%%%%%
\renewcommand*{\thefootnote}{\fnsymbol{footnote}}
\footnotetext[1]{%Author affiliations are listed at the end of this paper. \newline 
Corresponding author: \texttt{kjuliana@physics.upenn.edu}}
\footnotetext[2]{Corresponding author: \texttt{csanchez@ifae.es} }
\renewcommand*{\thefootnote}{\arabic{footnote}}
\setcounter{footnote}{0}

Since the discovery of cosmic acceleration, the nature of dark energy
has emerged as one of the most important open problems in
cosmology. Wide-field, large-volume galaxy surveys are promising
avenues to answer cosmological questions, since they provide multiple
probes of cosmology, such as Baryon Acoustic Oscillations (BAO), large
scale structure, weak lensing and cluster counts from a single
dataset. Moreover, some of these probes can be combined for greater
effect, since each is sensitive to their own combination of
cosmological parameters and systematic effects.  In this paper, we
will focus on combining the large scale angular clustering of galaxies
with measurements of the gravitational lensing produced by the large scale 
structure traced by the same galaxies, as observed in the Dark Energy Survey (DES).
%will focus on combining the large scale angular clustering of galaxies
%observed in the Dark Energy Survey (DES) with measurements of
%gravitational lensing by their host dark matter halos.
% to derive measurements of the dark energy density
%and equation of state.
 
Measurements of the large scale clustering of galaxies are among the
most mature probes of cosmology.  The positions of galaxies are seeded
by the distribution of dark matter on large scales and the manner in
which the growth of structure proceeds from gravitational collapse is
sensitive to the relative amounts of dark matter and energy in the
Universe. There is a long history of using large-volume galaxy surveys
for the purposes of constraining cosmology, including DES,
Sloan Digital Sky Survey (SDSS)~\citep{york00}, Hyper Suprime-Cam 
(HSC)~\citep{miyazaki12}, the Kilo-Degree Survey (KiDS)~\citep{dejong13, deJong15, kuijken15},
and the Canada France Hawaii Telescope Lensing Survey (CFHTLenS)~\citep{heymans12, erben13}.

Gravitational lensing, the deflection of light rays by massive
structures, provides a complementary method of probing the matter
distribution. Here we focus on galaxy-galaxy lensing~\citep{tvj1984, wb2000}, when
both the lenses and sources are galaxies. This involves correlating
the amount of distortion in the shapes of background galaxies with the
positions of foreground galaxies. The amount of distortion is
indicative of the strength of the gravitational potential of the lens
and therefore tells us about the amount of matter contained in the
lens plane.  Weak gravitational lensing produces two effects,
magnification of the source and shearing of its image, but this
analysis is only concerned with the latter. These have been used to
probe both cosmology~\citep{mandelbaum13,cacciato13,more15} and the
structure of dark matter halos and its connection to the galaxy distribution
and baryon content of the Universe~\citep{sjf2004,manetal06,manetal08,
  cvm2009, ltb12, gillis13, vvh14,  hudson15, sifon15, viola15, vanuitert16}.

Individual studies of large scale structure \citep{crocce15},
galaxy-galaxy lensing~\citep{clampitt16} and cosmic shear
\citep{becker15, DES15} using DES data as well as combined analyses
focusing on smaller scales~\citep{park15} have been presented
elsewhere. In this paper, we combine angular clustering and
galaxy-galaxy lensing to jointly estimate the large-scale galaxy bias
and matter clustering and constrain cosmological parameters.

The plan of the paper is as follows. Section~\ref{sec:theory} outlines
the theoretical framework for modelling the angular galaxy
correlation function and galaxy-galaxy lensing. Section~\ref{sec:data}
describes the galaxy sample used and the measurements from DES data, as well as the
covariance between the two probes. Our cosmology results are
summarized in Section~\ref{sec:cosmology} including 
constraints on a five-parameter $\Lambda$CDM (Cold Dark Matter) model
and a six-parameter $w$CDM model, where $w$, the dark energy equation of 
state parameter is also allowed to vary. 
We discuss the robustness of our results and our tests for systematic errors in 
Section~\ref{sec:testing}. Finally, we combine our analysis 
with other probes of cosmology and compare our results to previous results
in the literature in Section~\ref{sec:discussion}. Our conclusions
are presented in Section~\ref{sec:conclusion}.

%%%%%%%%%%%%%%%%%%%%%
\section{Theory}
\label{sec:theory}

We are interested in describing the angular clustering of galaxies, 
$w(\theta)$, and the tangential shear produced by  their host dark matter halos, 
$\gamt$, as a function of cosmology.  The angular correlation function, $w(\theta)$,
can be expressed in terms of the galaxy power spectrum as: 
\begin{eqnarray}
C({\ell}) &=& \frac{1}{c} \int d\chi \left(\frac{n_l(\chi) H(\chi)}{\chi}\right)^2 P_{gg}(\ell/\chi), \\
w(\theta) &=& \int \frac{\ell d\ell}{2\pi}C({\ell})J_0(\ell\theta), 
\label{eqn:wtheta}
\end{eqnarray}
where $P_{gg}$ is the galaxy auto power spectrum, $J_0$ is the Bessel function of order 0, 
$l$ is the angular wavenumber, $\chi$ is the comoving radial co-ordinate, $H(\chi)$ is the 
Hubble relation, $c$ is the speed of light, and $n_l(\chi)$ is  the number of galaxies  as a
function of radial distance from the observer, normalized such that
$\int^{\chi_{\rm max}}_{\chi_{\rm min}} n_l(\chi) \; d\chi = 1$.  
Note that Eq.~\ref{eqn:wtheta} uses the Limber approximation~\citep{limber53,kaiser92}, 
such that the radial distribution of galaxies, $n_l(\chi)$, is assumed
to be slowly varying over our redshift slice. We have also ignored
the contribution of redshift-space distortions to the angular clustering;
this is expected to be small due to the width of the redshift intervals
used; for the full expression, see~\citet{crocce15}.

%It is a measure of the
%amount of clustering integrated over the redshift interval of
%interest, and its sensitivity to cosmology is represented in the
%galaxy-galaxy power spectrum. 

The tangential shear is given by: 
\begin{equation}
\left<\gamma_t (\theta) \right> = 6\pi\Omega_m \int d\chi \,n_l(\chi) \frac{f(\chi)}{a(\chi)}\int dk \, k P_{g\delta}(k, \chi) J_2(k,\theta, \chi), 
\label{eqn:tangential_shear}
\end{equation}
where $f(\chi) = \int d\chi' n_s(\chi') \chi
(\chi-\chi')/\chi'$ is the lens efficiency, $a$ is the scale factor and $n_l(\chi)$ and
$n_s(\chi)$ are the selection functions of the lenses (foreground) and source
(background) galaxies respectively. The foreground galaxies supply the
gravitational potentials that lens the background galaxies. The
tangential shear is a measurement of the amount of distortion
introduced into the images of background galaxies from the
gravitational potentials of foreground galaxies as a function of
scale.  We will discuss the impact of photometric redshift (photo-$z$) errors on the lens and
source distributions and propagate these to the measured cosmological
constraints in Section~\ref{sec:testing}.

The combination of these two probes has been extensively discussed in
the literature \citep{baldauf10, yoo12, mandelbaum13, park15} and
provide another means by which we can mine the rich, well calibrated
DES-SV dataset. Unlike \citet{park15}, we restrict our modelling to
sufficiently large scales such that we are not sensitive to how
galaxies populate individual halos, i.e. Halo Occupation Distribution
(HOD) modelling is unnecessary.  On these scales, we are only
concerned with correlations between galaxies that reside in different
halos (the 2-halo term of the power spectrum), and we can relate the
matter power spectrum, $P_{\delta\delta}$, to the galaxy power
spectrum $P_{gg}$ and galaxy-dark matter cross-power spectrum
$P_{g\delta}$ via the following relationships:
\begin{eqnarray}
P_{gg} (k) \approx b_{\rm g}^2 P_{\delta\delta}(k),  \label{eqn:pow_gg} \\
P_{g\delta}(k) \approx b_{\rm g} r P_{\delta\delta} (k), \label{eqn:pow_gd}
\end{eqnarray}
where $b_{g}$ is the linear bias that relates the clustering of
galaxies to that of dark matter and $r$ is the cross-correlation
coefficient that captures the stochasticity between the clustering of
dark matter and the clustering of galaxies; see for example~\citet{seljak00,guzik01}.

%, and is
%defined as:
%\begin{equation}
%r = \frac{P_{g\delta}}{\sqrt{P_{\delta\delta}P_{gg}}}
%\label{eqn:r}
%\end{equation}

The measurement of $w(\theta)$ depends on $b_{\rm g}^2
P_{\delta\delta}$, while the tangential shear, $\gamt$, depends on
$b_{\rm g} P_{\delta\delta}$ if $r=1$, a reasonable approximation on
the large scales we use in this work (we allow for and marginalize
over possible stochasticity through our non-linear bias modelling; see
Section \ref{sec:nonlinearbias}). The measurements of $w(\theta)$ and
$\gamt$ in combination allow us to estimate both the clustering
amplitude and the linear galaxy bias, thus enabling us to obtain
useful cosmological information.

\subsection{Non-linear bias model}\label{sec:nonlinearbias}
The assumption of linear bias in Eqs.~(\ref{eqn:pow_gg})~and~(\ref{eqn:pow_gd}) 
is expected to break down at small scales. In order to account for this effect, we
use the non-linear biasing scheme of~\citet{mcdonald06}, where the
galaxy over-density, $\delta_g$, is written as
\begin{equation}
\delta_g = \epsilon + b_1\delta + b_2\delta^2 + {\mr {next \; leading \; order \; bias \; terms}}, 
\end{equation}
where $b_1$ is the usual linear bias, $b_2$ is the next leading order
bias term and $\epsilon$ is the shot noise. The bias parameters, $b_1$
and $b_2$ are not known a priori and become free parameters to be
constrained during the analysis.  Under this perturbation
theory scheme, the galaxy-dark matter and galaxy-galaxy power
spectra become
\begin{eqnarray}
P_{g\delta} & =& b_1 P_{\delta\delta} + b_2 A(k),  \\
P_{gg} &=& b_1^2 P_{\delta\delta} + b_1 b_2 A(k) + b_2 B(k) + N, 
\label{eqn:nonlinpow}
\end{eqnarray}
where $N$ is the shot noise and $A(k)$ and $B(k)$ can be calculated using standard 
perturbation theory as follows:
\begin{eqnarray}
A(k) = \int \frac{d^3q}{(2\pi)^3} F_2({\bf q, k-q}) P_{\delta\delta}(q) P_{\delta\delta}(|{\bf k-q}|), \\
B(k) = \int \frac{d^3q}{(2\pi)^3} P_{\delta\delta}(q) P_{\delta\delta}(|{\bf k-q}|), 
\label{eqn:pt}
\end{eqnarray}
where $F_2({\bf k_1, k_2}) = \frac{5}{7}\frac{({\bf k_1} + {\bf k_2})
  \cdot {\bf k_2}}{k_1^2} + \frac{1}{7} ({\bf k_1} + {\bf k_2})^2
\frac{{\bf k_1} \cdot {\bf k_2}}{k_1^2k_2^2}$.  Note that this
non-linear biasing scheme generates departures from $r = 1$ as $r \approx 1-1/4(b_2/b_1)^2
\xi_{gg}$, where $\xi_{gg}$ is the correlation function.  
As such we do not include an additional free parameter for the 
cross-correlation coefficient. We found that for reasonable values of the shot noise, $N$,
given the density of our galaxy sample, has a less than 5\% effect on $\wtheta$ on scales below our regime
of interest ($<20'$) and so have ignored this term for the remainder
of our analysis. We do, however, include an additional additive
constant term in configuration space as discussed in
Section~\ref{sec:testing}. This term mainly alters the large scale
clustering to allow for possible systematics coming from 
observational effects (see Section~\ref{sec:lss_systematics}). 

We investigate the inclusion of the next order biasing term in
Section~\ref{sec:scale}, in which we vary both the lower limit on
the angular scale cutoff and the modelling of non-linear bias.

%%%%%%%%%%%%%%%4%%%%%%
\section{Data and measurements}
\label{sec:data}

\begin{figure}
\centering
\resizebox{90mm}{!}{\includegraphics{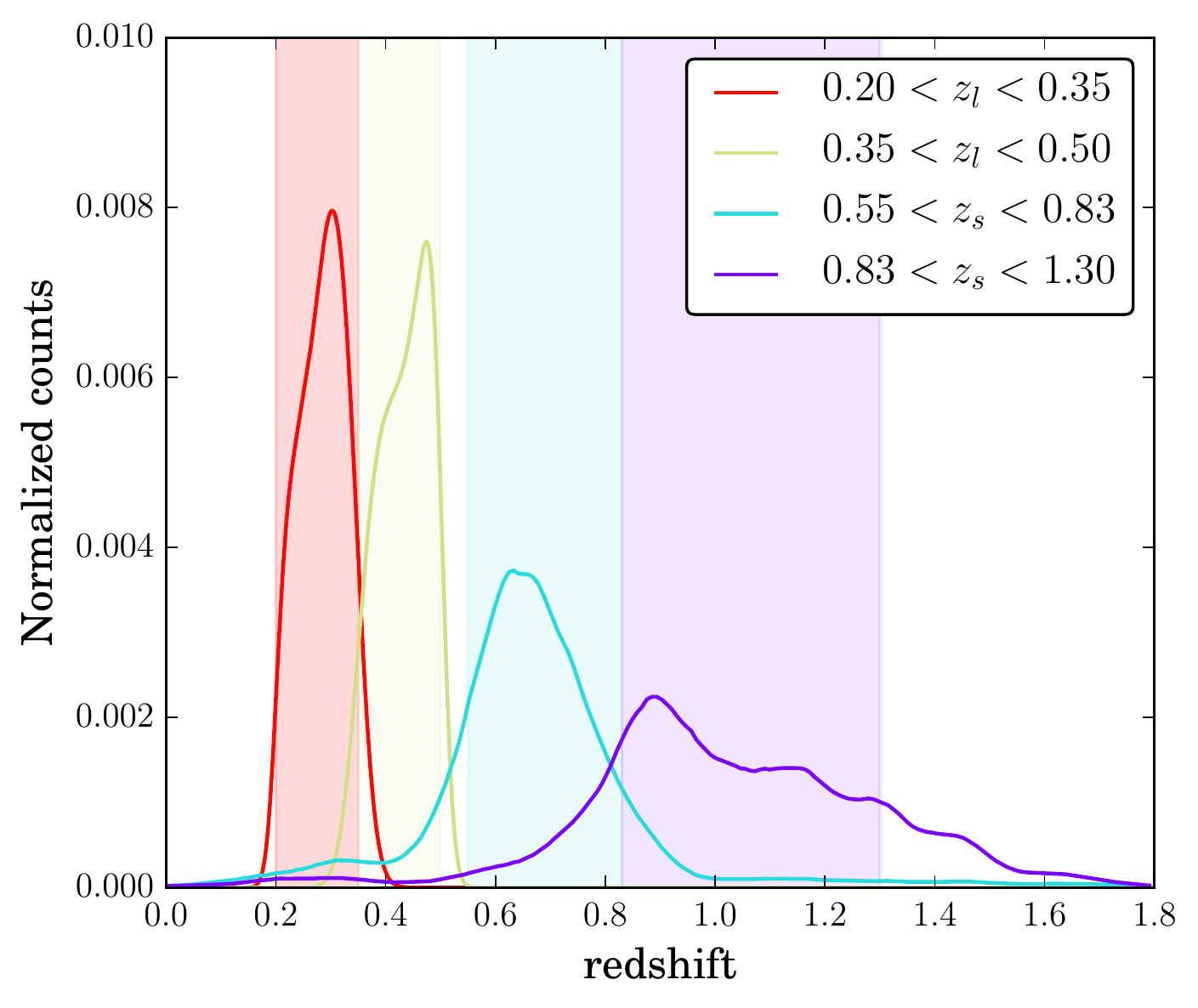}}
\caption{ Redshift distributions of the four galaxy samples used in
  this work. Red and yellow curves correspond to the two \redmagic{}
  lens bins while cyan and purple curves correspond to the two source
  bins in the fiducial configuration (\texttt{ngmix} shears, SkyNet
  photo-$z$'s).}
\label{fig:nzs}
\end{figure}

The Dark Energy Survey (DES) is an ongoing photometric survey that
aims to cover 5000 sq.~deg.~of the southern sky in five photometric
filters, $grizY$, to a depth of $i \sim 24$ over a five year
observational program using the Dark Energy Camera (DECam,~\citet{flaugher15}) 
on the 4m Blanco Telescope at the Cerro Tololo
Inter-American Observatory (CTIO) in Chile.  In this analysis, we will be utilizing
DES-SV (Science Verification) data, in particular a contiguous $\sim$
139 sq.~deg.~patch known as the SPT-E region (because of its overlap
with the South Pole Telescope survey footprint). This is only a small
($\sim$ 3\%) subset of the expected eventual sky coverage of DES, but
observations in all five filters have been performed at full depth, 
although substantial depth variations are present (see e.g. \citealt{leistedt15}), 
mainly due to weather and early DECam operational challenges.
The DES-SV data have been used for constraining cosmology in this work, 
but a rich variety of science cases are possible with this data sample 
(see \citet{DES16} and references therein).

The lens galaxy sample used in this work is a subset of the DES-SV
galaxies selected by \redmagic{}\footnote{\texttt{https://des.ncsa.illinois.edu/releases/sva1}}~\citep{rozo15b}, which is an algorithm
designed to define a sample of Luminous Red Galaxies (LRGs) by
%%BJ: Is that how Rozo et al describe redmagic -- a sample of LRGs? if not, let's replace by "bright red galaxies"
minimizing the photo-$z$ uncertainty associated with the
sample.  It selects galaxies based on how well they fit a
red sequence template, as described by their goodness-of-fit,
$\chi^2$. The red sequence template is calibrated using redMaPPer
\citep{rykoff14,rozo15a} and a subset of galaxies with
spectroscopically verified redshifts.  The cutoff in the goodness of
fit, $\chi_{\rm cut}^2$, is imposed as a function of redshift and 
adjusted such that a constant comoving number density of galaxies is
maintained, since red galaxies are expected to be passively evolving.
The \redmagic{} photo-$z$'s show excellent performance,
with a median photo-$z$ bias, $(z_{\rm spec}-z_{\rm phot})$, of 0.005
and scatter, $\sigma_z/(1+z)$, of 0.017. Equally important, their errors are  very
well characterized, enabling the redshift distribution of a sample,
$N(z)$, to be determined by stacking each galaxy's Gaussian
redshift probability distribution function (see
\citealt{rozo15b} for more details).

%For this reason, we have chosen to use the high density \redmagic{} catalog for 
%our lenses and angular clustering sample. 
%paragraph on the source sample (ngmix and im3shape (fiducial ngmix), photo-$z$'s Chris paper (fiducial SkyNet),etc)

The galaxy shape catalogs used in this work were presented in
\citet{jarvis15}, and they have been used in several previous analyses
\citep{vikram15, becker15, DES15, gruen15, clampitt16}. Two different 
catalogs exist corresponding to the
\texttt{ngmix}~\footnote{{\texttt{https://github.com/esheldon/ngmix}}}
\citep{sheldon14} and
\texttt{im3shape}~\footnote{\texttt{https://bitbucket.org/joezuntz/im3shape}}
\citep{2013MNRAS.434.1604Z} shear pipelines, both producing model
fitting shape measurements to a subset of DES-SV galaxies. The two
catalogs differ in their approach to modelling the intrinsic galaxy
shape ({\ngmix} uses a Gaussian mixture model to approximate an
exponential disk galaxy profile while {\imshape} determines the maximum
likelihood for fitting a bulge and/or disk profile) and also in the
number of filters used ({\ngmix} uses $riz$ bands while {\imshape} only uses $r$
band).  This results in the {\ngmix} catalog containing more sources
than {\imshape} ($\sim$6.9 galaxies per arcmin$^2$ vs.~$\sim $4.2
galaxies per arcmin$^2$). More details about the pipelines and an
extensive set of null and systematics tests can be found in
\citet{jarvis15}. The photo-$z$ distributions of the galaxies in the shear catalogs 
were studied in detail in \citet{bonnett15}, using 4 different photo-$z$
codes that performed well in a previous more extensive photo-$z$ code comparison \citep{sanchez14}. 
The four methods are SkyNet \citep{graff14,bonnett15b}, ANNz2 \citep{sadeh15},
TPZ \citep{carrascokind13} and BPZ \citep{2000ApJ...536..571B}.
The first three methods are training-based, and the last is a widely used template-based code.
Details about their training or calibration procedures and about the validation 
against spectroscopic data can be found in \citet{bonnett15}.

\begin{figure}
\centering
\resizebox{90mm}{!}{\includegraphics{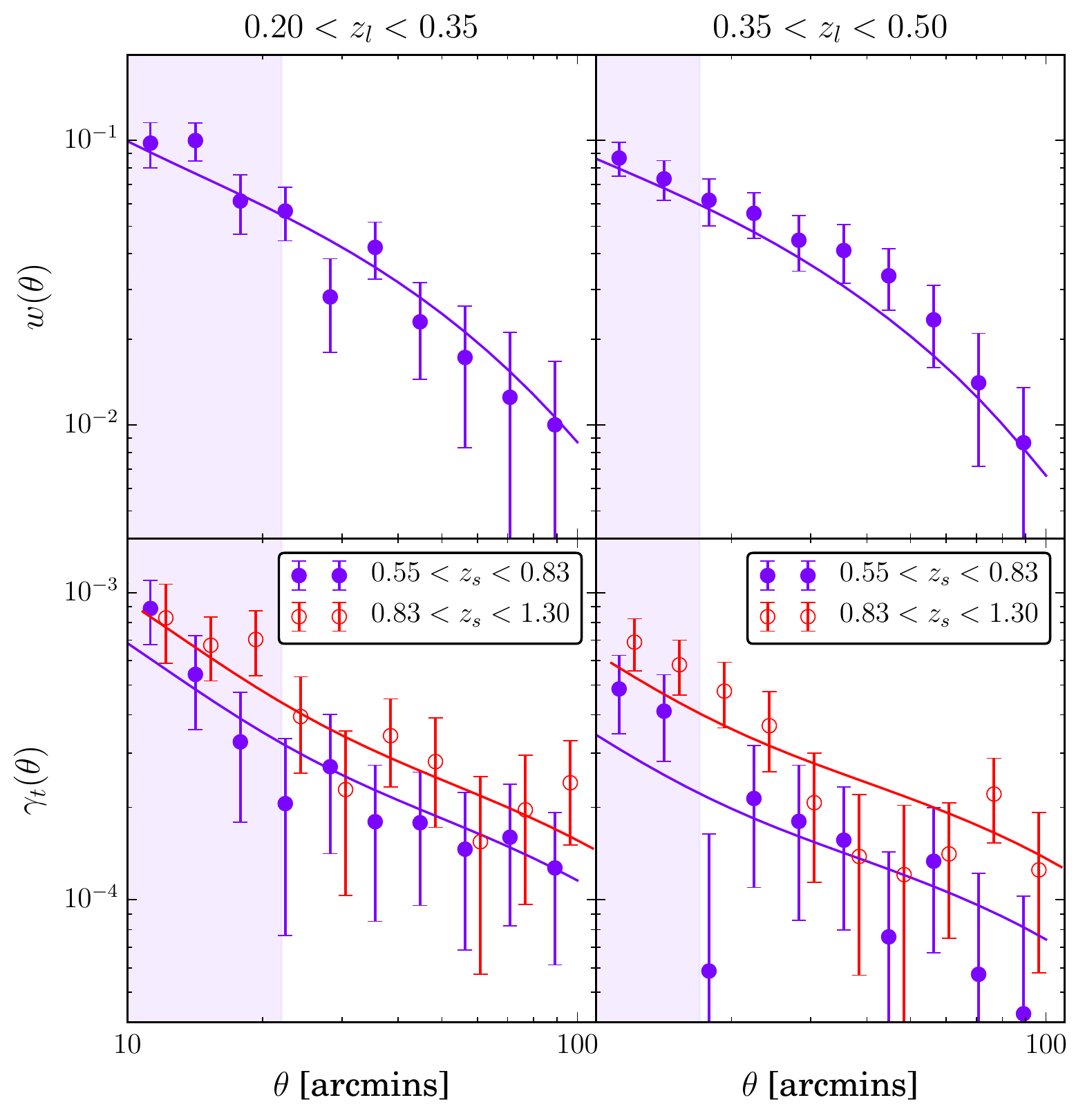}}
\caption{Angular galaxy clustering and galaxy-galaxy lensing
  measurements used in this work. For the two lens bins (left and
  right columns), we show the clustering measurements (upper row) and
  the galaxy-galaxy lensing measurements (lower row) for the two
  source bins, with error bars coming from jackknife resampling. 
  The shaded region shows excluded scales in the fiducial analysis, explored in Section~\ref{sec:scale}. 
  The predictions for the best fitting curves presented in Section~\ref{sec:cosmology} 
  are shown as the solid curves in each panel. The goodness-of-fit, as measured
  by the $\chi^2$ value is 6 (3.5) for 12 (9) degrees of freedom for the high$-z$ (low-$z$) bin.}
\label{fig:measurements}
\end{figure}

In this paper we use the \texttt{ngmix} shear catalog and SkyNet 
photo-$z$'s for the fiducial results, but we will test the robustness of
our results with the {\imshape} shear catalog  as well as using the
source distributions derived from the other photo-$z$ algorithms
in the analysis.

%%%%%%%%%%%%%%%%%%%%%
\subsection{Measurements}\label{sec:measurements}

%In this work, 
%we use the combination of angular galaxy clustering and
%galaxy-galaxy lensing on large scales to constrain cosmology. We use
We use two lens bins, selected using \redmagic{} photo-$z$'s: 
$0.20 < z < 0.35$ and $0.35 < z < 0.50$, and two source bins, 
 selected using SkyNet photo-$z$'s: $0.55 < z < 0.83$ and
$0.83 < z < 1.30$. The same lens photo-$z$ bins are analyzed in \citet{clampitt16} while the source photo-$z$ bins are studied in detail in~\citet{bonnett15}
and used for cosmology in~\citet{DES15}. Individual
analyses involving $\gamt$ and $\wtheta$ with DES-SV have been
presented in~\citet{clampitt16} and \citet{crocce15}, respectively. 
Figure~\ref{fig:nzs} shows the redshift distributions for the lens and source
bins utilized in this analysis. For each lens bin, we measure the
galaxy clustering and the galaxy-galaxy lensing signals using the
estimators defined next. The correlation functions have been estimated 
 using the code \verb+TreeCorr+\footnote{\texttt{https://github.com/rmjarvis/TreeCorr}}
~\citep{jarvis04}. 
%For each lens bin, we measure the clustering signal and the two galaxy-galaxy lensing signals (corresponding to the two source bins) using the estimators defined next.  

\subsubsection{Angular Clustering -- $w(\theta)$}
On the galaxy clustering side, we compute the angular correlation
function for each redshift bin using the minimum variance estimator of
\citet{landy93},
\begin{equation}
w(\theta) = \dfrac{\textrm{DD}-2\textrm{DR}+\textrm{RR}}{\textrm{RR}} \, ,
\end{equation}
where $\theta$ is the angular separation in the sky, and DD, DR and RR
are data-data, data-random and random-random pairs of galaxies, with
data and random galaxies having the exact same geometry in the sky.
The resulting measurement is shown in Fig.~\ref{fig:measurements}.
The clustering amplitude falls from $\sim 10^{-1}$ to $10^{-2}$ over
the range $\theta = 10-100$ arcminutes. Only scales $\sim 20$
arcminutes and above will be used in the cosmology fits (see
Sec.~\ref{sec:scale} for details). The details of the calculation of
the error or covariance matrix for $\wtheta$ will be presented 
in Section~\ref{sec:covariances}.

\begin{figure*}
\centering
\resizebox{190mm}{!}{\includegraphics{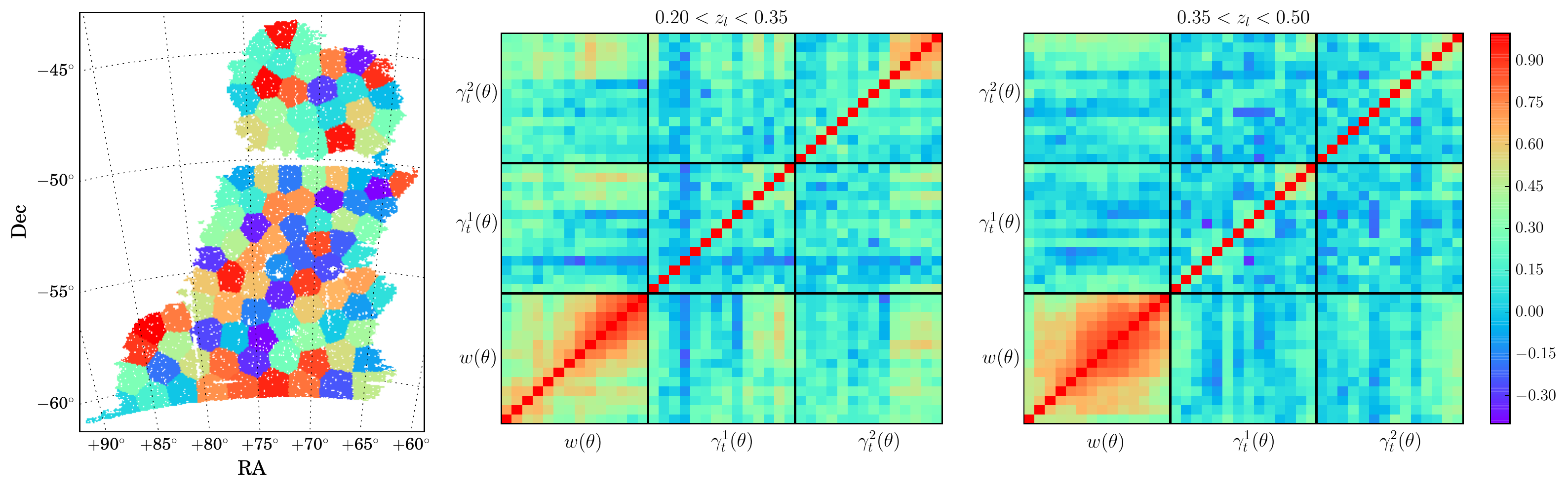}}
\caption{(\textit{left panel}): DES-SV SPT-E footprint and an example of
  the \textit{kmeans} jackknife regions used to compute the covariance
  matrices used in this work. (\textit{center panel}): For the first
  lens bin, the joint jackknife correlation matrix for  
  $w(\theta)$ and $\gamt$ for the two source bins. For each submatrix
  of the joint correlation matrix, the angular scale ranges from 4 to 100
  arcminutes in logarithmic bins. (\textit{right panel}): Same as the 
  \textit{center panel}, for the second lens bin.}
\label{fig:jk_covs}
\end{figure*}

\subsubsection{Tangential Shear -- $\gamma_t(\theta)$}
On the lensing side, the  observable is the tangential
shear, i.e., the shear of the source galaxy which is perpendicular to the
projected line joining the lens and source galaxies. For a given lens-source
pair ($j$) this is given by
\begin{equation}
\gamma_{t,j} = -\gamma_{1,j} \cos(2\phi_j) -\gamma_{2,j} \sin(2\phi_j)
\label{eqn:tangentialshear}
\end{equation}
where $\gamma_{1,j}$ and $\gamma_{2,j}$ are the two components of
shear measured with respect to a cartesian coordinate system with origin in the lens galaxy, and
$\phi_j$ is the position angle of the source galaxy with respect to
the horizontal axis of the cartesian coordinate system. Since the
intrinsic ellipticity of individual source galaxies is much larger
than the weak lensing shear, it is necessary to average over many such
lens-source pairs. For our measurements, we compute the average in
angular separation bins, $\theta$, so that
\begin{equation}
\ave{\gamma_t (\theta)} = \frac{\sum_j \omega_j \gamma_{t,j}}{\sum_j \omega_j} \, ,
\end{equation}
where the tangential shear for each lens-source pair, $j$, is weighted by 
a factor $\omega_j$ as follows: \be \omega_j = \dfrac{1}{\sigma_{\rm
    shape}^2 + \sigma_{{\rm m},j}^2} \, , \ee where $\sigma_{\rm
  shape}$ is the shape noise intrinsic to each background galaxy, and
$\sigma_{{\rm m},j}$ is the error derived from the shape measurement.
The weights $\omega_j$ corresponding to the shear catalogs used in this
work are computed and described in~\citet{jarvis15}. 
%A small constant blinding 
%factor is applied to the ellipticities to avoid experimenter bias. The blinding factor 
%shifts contours in the $\Omega_m - \sigma_8$ plane away from expected values, 
%such that these results can not be used to calibrate the analysis pipeline.
In order to correct for possible geometric and additive shear systematic effects,
we compute the tangential shear around random lenses and subtract this
from the galaxy lensing signal (as in~\citet{clampitt16}). The
result is shown in the lower panels of Fig.~\ref{fig:measurements},
over the same range of scales as for $w(\theta)$. For each lens bin we
show the tangential shear using the two source bins.
%As expected, for fixed lens bin the shear signal
%increases with source redshift (see the lensing weight in
%Eq.~(\ref{eq:sigma_crit})). 
%The magnitude of the difference helps
%constrain cosmology, which will be exploited in the following
%sections.

\subsection{Covariances}\label{sec:covariances}

Our measurements of $\wtheta$ and $\gamt$ are correlated across angular and source redshift bins. 
The joint covariance for all the measurements corresponding to each
lens redshift bin is estimated from jackknife (JK) resampling,
using the following expression \citep{norberg09}:
\begin{equation}
C(x_i,x_j) = \frac{(N_{\rm JK}-1)}{N_{\rm JK}}\sum_{k=1}^{N_{\rm JK}} (x_i^k - \bar{x}_i)(x_j^k - \bar{x}_j),
\end{equation}
where the complete sample is split into a total of $N_{\rm JK}$ groups, $x^k_i$
is a measure of the statistic of interest in the $i$-th bin using all JK regions 
excepting the $k$-th sample, and $\bar{x}_i$ is the mean of $N_{\rm JK}$
resamplings. Jackknife regions are obtained using the \textit{kmeans}
algorithm\footnote{\texttt{https://github.com/esheldon/kmeans\_radec}}
run on a homogeneous random points catalog and, then, all catalogs
(lenses, sources and random points) are split in $N=100$ JK
samples. \textit{kmeans} is a clustering algorithm that subdivides $n$
observations into $N$ clusters (see Appendix B in \citealt{suchyta16} for details). 
By applying it to a uniform random catalog with the same sky coverage as DES-SV, 
we define regions that are well suited for JK subsampling.
The left panel in Fig.~\ref{fig:jk_covs} shows our JK patches created by the
\textit{kmeans} algorithm.  The resulting covariance matrices for both
lens bins are also shown in Fig.~\ref{fig:jk_covs} (center and right panels).  The covariance is
strongest between points within the $w(\theta)$ data vector.  Note
that: (i) we do not jointly fit both lens bins in the fiducial case so no covariances
between lens bins are shown, and, (ii) when performing cosmology fits
with the lower (higher) lens bin we only use 21 (24) data points (see
Sec.~\ref{sec:scale}).

The JK covariance matrices shown in Fig.~\ref{fig:jk_covs} contain a 
non-negligible level of noise. \citet{hartlap07} showed that the inverse 
of an unbiased but noisy estimator of the covariance matrix is actually 
\textit{not} an unbiased estimator of the inverse covariance matrix. 
Therefore, when using a JK covariance matrix, a correction factor 
of $(N_{\rm JK}-N_{\rm bins}-2)/(N_{\rm JK}-1)$ should be applied to the 
inverse covariance, where $N_{\rm JK}$ is the 
number of jackknife regions and $N_{\rm bins}$ is the number of
measurements \citep{hartlap07}. We include this correction factor in all our cosmology results.

The performance of JK covariances in DES-SV has been studied
separately for galaxy clustering and galaxy-galaxy lensing in
\citet{crocce15,giannantonio15} and \citet{clampitt16},
respectively. There we generally find good agreement between true
covariances from simulations or theory and the JK estimates,
especially at small scales. At large scales the comparison points to
an overestimation of the covariance by the JK method in the lensing
case.
%, although differences are rather small (less than a factor of 2) and go in the conservative direction. 
%In this work we also estimate the cross-covariance between galaxy
%clustering and galaxy-galaxy lensing, a contribution that was
%neglected in related previous analyses (e.g. \citet{mandelbaum13}). We
%find a positive correlation among all clustering scales and large
%galaxy-galaxy lensing scales -- the regime where the lensing errors
%are no longer dominated by shape noise.
In this work we also estimate the cross-covariance between galaxy clustering and galaxy-galaxy lensing, for which we find a small positive correlation among all clustering scales and large galaxy-galaxy lensing scales -- the regime where the lensing errors are no longer dominated by shape noise. This is consistent with related previous work like \citet{mandelbaum13}, where they were able to neglect this contribution due to their different noise properties. As a check on the amount of covariance between probes, Fig.~\ref{fig:constraints_table} also shows the result of ignoring the cross-covariance on the constraints on $\Omega_m$ and $\sigma_8$. The derived cosmology shows little deviation from our fiducial results and we find that our
constraints are only minimally stronger on $\sigma_8$ (by about 3\%) and weaker on $\Omega_m$ (also $\sim$3\%) with
a 2\% improvement on S$_8 = \sigma_8(\Omega_m/0.3)^{0.16}$. This shows that the impact of the correlation between probes is subdominant to the covariance within the same probe. 

\begin{figure*} 
\centering
\begin{minipage}{.49\textwidth}
%\vspace{-0.3cm}
   \centering
   \includegraphics[width=\linewidth]{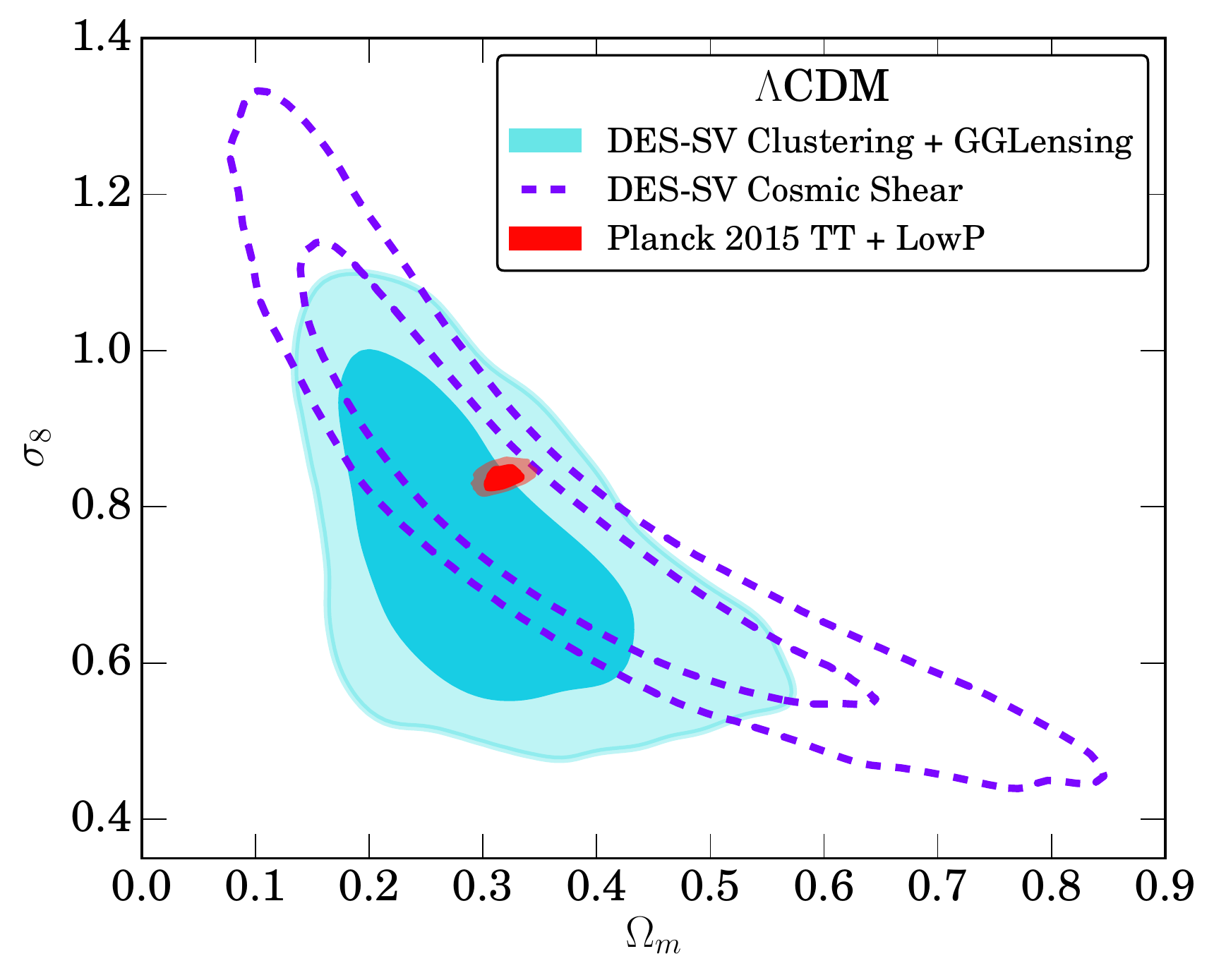} 
   %\resizebox{90mm}{!}{\includegraphics{omega_m_sigma_8_high_z_shear_2pt_planck2015.pdf}}
   %\includegraphics[width=\linewidth]{sigma_8_omega_m_high_z_w_shear_2pt.png} 
   \caption{Constraints on $\Omega_m$ and $\sigma_8$ using DES-SV
       Cosmic Shear (dashed purple), DES-SV $w(\theta) \; \times\; \gamt $
       (this work, filled blue) and Planck 2015 using a combination of
       temperature and polarization data (TT+lowP, filled red). In each case,
       a flat $\Lambda$CDM model is used.}
   \label{fig:comparison_shear_2pt}
\end{minipage}
\hfill
\begin{minipage}{.49\textwidth}
%\vspace{-0.3cm}
   \includegraphics[width=\linewidth]{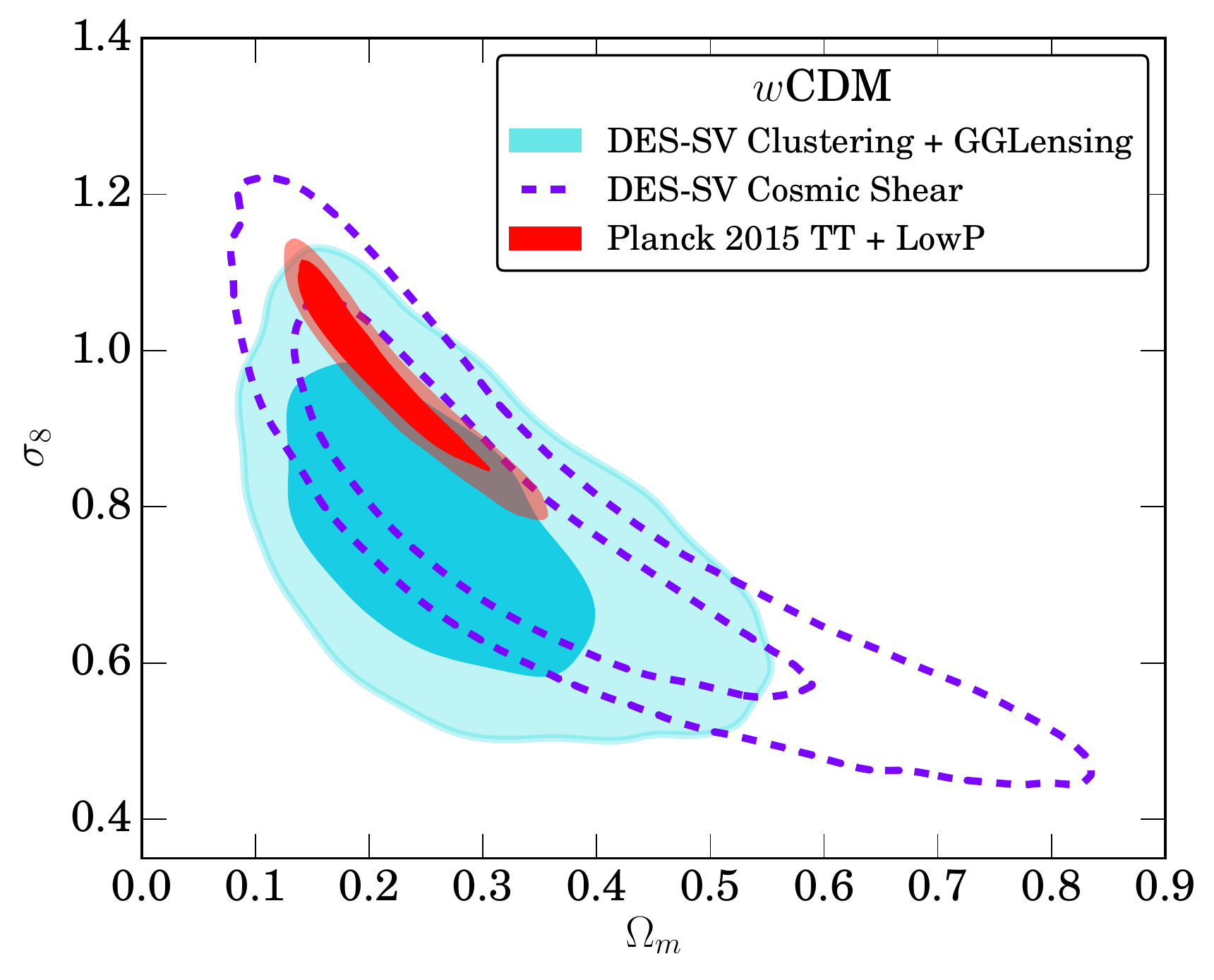} 
   %\resizebox{90mm}{!}{\includegraphics{omega_m_sigma_8_high_z_shear_2pt_planck2015_wcdm.pdf}}
   %\includegraphics[width=\linewidth]{sigma_8_omega_m_high_z_w_shear_2pt.png} 
   \caption{Constraints on $\Omega_m$ and $\sigma_8$ assuming a
       $w$CDM model using DES-SV Cosmic Shear (dashed purple), DES-SV 
       $w(\theta) \; \times\; \gamt $ (this work, blue) and Planck 2015 using 
       temperature and polarization data (TT+lowP, red). }

   \label{fig:wCDM}

\end{minipage}
\end{figure*}
%%%%%%%%%%%%%%%%%

\section{Fiducial Cosmological Constraints}
\label{sec:cosmology}

\begin{table*}
	\centering
	\begin{tabular}{@{}ccl@{}}
		\toprule 
		 Parameter & Prior range & \\
		\midrule
		$\Omega_m$    & 0.1 -- 0.8      & Normalized matter density \\
		$\Omega_b$    & 0.04 -- 0.05    & Normalized baryon density \\
		$\sigma_8$    & 0.4 -- 1.2      & Amplitude of clustering (8 h$^{-1}$Mpc top hat) \\
		$n_s$         & 0.85 -- 1.05    & Power spectrum tilt \\
		$w$           & -5 -- -0.33   & Equation of state parameter \\
		$h$           & 0.5 -- 1.0    & Hubble parameter (H$_0$ = 100$h$) \\
		$\tau$        & 0.04 -- 0.12    & Optical depth  \\
		\midrule
		$b_1$         & 1.0 -- 2.2      & Linear galaxy bias  \\
		$b_2$         & -1.5 -- 1.5     & Next order bias parameter \\
		$\beta_i$     & -0.3 -- 0.3   & Shift in photo-$z$ distribution (per source bin)\\
		$m_i$         & -0.2 -- 0.2   & Shear multiplicative bias (per source bin)\\ 
		$m_{IA}$    & -0.3 -- 0.35   & Intrinsic alignment amplitude (low-z source bin only)\\
		$\alpha$& -5 -- -1        & Additive constant $w(\theta) \rightarrow w(\theta) + 10^{\alpha}$ \\
		\bottomrule
	\end{tabular}

	%\caption{}
	
	\caption{
	Parameters and their corresponding priors used in this work. Not all parameters are
          allowed to vary in every analysis. Nuisance parameters are contained in the lower half of the
          table. When choosing a prior range on cosmological parameters, we allowed a sufficiently
          wide range to contain all of the 2-$\sigma$ posterior on $\Omega_m$, $\sigma_8$, $n_s$, $w$ 
          and $h$, with Planck priors on $\Omega_b$, for which we have less sensitivity. For the systematic parameters,
          our choice of prior range is informed from previous DES analyses that studied the effect of shear 
          calibration~\citep{jarvis15}, photo-$z$ distributions~\citep{bonnett15}, and intrinsic alignment 
          contamination~\citep{clampitt16, DES15} on the SV catalogues. The prior on the bias parameters
          were taken from studies of the \redmagic{} mock catalog (see Section~\ref{sec:scale} for details).
          In addition to the prior range on the nuisance parameters for the shear calibration and photo-$z$
          bias, there is a Gaussian prior centered around zero of width 0.5, as explained in the text.}
          \label{tab:parameters}
\end{table*}

\begin{table*}

   \centering
   %\topcaption{Table captions are better up top} % requires the topcapt package
   \begin{tabular}{@{} lcccccccc@{}} % Column formatting, @{} suppresses leading/trailing space
      \toprule
      Probes            & $z$                & $\sigma_8$           & $\Omega_m$     & S$_8\equiv\sigma_8(\Omega_m/0.3)^\alpha$& $\alpha$    & $b_1$                             & $w_0$ \\
      \midrule
      DES               & 0.2 $< z <$ 0.35 &  0.73 $\pm$ 0.12  &   0.46 $\pm$ 0.12  & 0.77 $\pm$ 0.11 & 0.15   &
      1.60 $\pm$  0.31   &   -1       \\
      
      DES               & 0.2 $< z <$ 0.35 &  0.74 $\pm$ 0.13  &   0.41 $\pm$ 0.14  & 0.77 $\pm$ 0.10 & 0.17   &
      1.73$\pm$  0.29   &   -2.5 $\pm$ 1.26      \\
      
      DES               & 0.35 $< z <$ 0.5 &  0.74 $\pm$ 0.13  &   0.31 $\pm$ 0.09  &  0.74 $\pm$ 0.12 & 0.16  & 
      1.64 $\pm$  0.30   &   -1        \\
      
      DES               & 0.35 $< z <$ 0.5 &  0.77 $\pm$ 0.12  &   0.28 $\pm$ 0.10  &  0.75 $\pm$ 0.11  & 0.13  & 
      1.71 $\pm$  0.28   &   -2.03$\pm$ 1.19      \\
      
      DES               & 0.2 $< z <$ 0.5 &  0.76 $\pm$ 0.10  &   0.36 $\pm$ 0.09  &  0.78 $\pm$ 0.09  & 0.21  & 
      1.52 $\pm$  0.28  &   -1      \\
      
                            &                           &                             &                                &                              &          & 
     1.60 $\pm$ 0.27   &         \\                   
      
      Planck                 &                       &  0.83 $\pm$ 0.01  &   0.32 $\pm$ 0.01  &  0.82 $\pm$ 0.02 & -0.49 &
                                     & -1 \\
       
      Planck                 &                       &  0.98$^{+0.11}_{-0.06}$ & 0.21$^{+0.02}_{-0.07}$ &  1.21$\pm$0.27 &  -0.6 &
                                  &-1.54$^{+0.20}_{-0.40}$ \\
      
      BAO + SN + H0   &                            &                                 &   0.33 $\pm$ 0.02   &                       &         &
                                  & -1.07 $\pm$ 0.06 \\
                                  
      BAO + SN + H0  + DES &  0.35 $< z <$ 0.5 &  0.71$\pm$0.1&   0.32 $\pm$ 0.02   &  0.71 $\pm$ 0.1 &  0.01 &
                                  & -1.05 $\pm$ 0.07 \\

      DES + Planck      & 0.2 $< z <$ 0.35 &  0.84 $\pm$ 0.01  &   0.35 $\pm$ 0.01  &  0.76 $\pm$ 0.02 & -0.71 &
      1.30 $\pm$  0.13   &  -1     \\
      
      DES + Planck      & 0.2 $< z <$ 0.35 &  0.89 $\pm$ 0.03  &   0.32  $\pm$ 0.02 & 0.84 $\pm$ 0.06   & -0.76 &
      1.25 $\pm$  0.13   &   -1.16 $\pm$ 0.09      \\
      
      DES + Planck      & 0.35 $< z <$ 0.5 &  0.84 $\pm$ 0.01  &   0.35 $\pm$ 0.01  &  0.76 $\pm$ 0.02  & -0.71 &
      1.41 $\pm$  0.17   &   -1    \\     
      
      DES + Planck      & 0.35 $< z <$ 0.5 &  0.88 $\pm$ 0.03  &   0.32 $\pm$ 0.02  &  0.84 $\pm$ 0.06   & -0.75 &
      1.36 $\pm$  0.14   &   -1.14 $\pm$ 0.09      \\   
      
      DES + Planck + & 0.35 $< z <$ 0.5 &  0.86 $\pm$ 0.02 & 0.31 $\pm$ 0.01 & 0.84 $\pm$ 0.03    &-0.81  &
      1.74 $\pm$ 0.28 & -1.09 $\pm$ 0.05 \\
       BAO + SN + H0 & & & & & & & \\
      \bottomrule
         %\label{tab:results}
   \end{tabular}
   \caption{
   Marginalized mean cosmological parameters (and 1-$\sigma$ errors) measured from the posterior distribution of a joint analysis of angular clustering and galaxy-galaxy lensing. Results for DES-SV data alone and in combination with Planck and
   external data (BAO, SN1a, H0) are shown for the two lens redshift bins both separately and combined. (Note that the biases
   are quoted separately: $b_1 = 1.52 \pm 0.28 $ for 0.2 $<z<$ 0.35 and $b_1 = 1.60 \pm 0.27$ for 0.35 $< z <$ 0.5). Not shown are the additional cosmological parameters that we have marginalized, $\{n_s,\Omega_b, h_0\}$ as well as our standard set of nuisance parameters. Also quoted are the mean values and 1-$\sigma$ errors given by Planck (TT+lowP) and external data alone. }
\label{tab:results}
\end{table*}

In this section we present our fiducial DES-SV cosmological
constraints from a joint analysis of clustering and galaxy-galaxy lensing.
The  data vector consists of $\wtheta$ and the two
$\gamt$ measurements for the $0.35<z<0.5$ \redmagic{} bin
(see Fig.~\ref{fig:measurements}), over angular scales of 17-100 arcminutes.
We chose this lens bin as our fiducial, as we estimate 
greater contamination from systematic errors, on both the clustering and lensing side, 
for the $0.2<z<0.35$ \redmagic{} bin (see Section~\ref{sec:lss_systematics} and \citet{clampitt16}).
%(For a $\Lambda$CDM cosmology with $\Omega_m = 0.31$, $\Omega_\Lambda = 0.69$, $w = -1$ and $h = 0.67$, the chosen minimum angular scale corresponds to $\sim 5$ Mpc/h. The choice of scales and its impact on the results will be explored in Section~\ref{sec:testing}.)
To compute the model we use CAMB \citep{lewis00, howlett12} and
Halofit \citep{smith03, takahashi12} for the linear and non-linear
matter power spectra, respectively. Because the accuracy of Halofit can
be confirmed only to $\sim$5\% for certain $\Lambda$CDM models, we
have checked that using the Cosmic Emulator, a more precise modelling 
scheme for the nonlinear dark matter power spectrum (1\% to $k = 1$ Mpc$^{-1}$,
\citealt{lawrence10}) would only affect our results at the level of $\sim$5\% 
down to $10'$. We use the CosmoSIS package\footnote{\texttt{https://bitbucket.org/joezuntz/cosmosis}}
\citep{zuntz15} as our analysis pipeline and explore the joint
posterior distribution of our cosmological (and nuisance) parameters
using the multi-nest MCMC algorithm of~\citet{feroz09}, with a tolerance
parameter of 0.5 and an efficiency parameter of 0.8. Our cosmological 
parameters and priors are summarized in Table~\ref{tab:parameters} and described 
in greater detail next in this section.

In the fiducial case, we have included two nuisance parameters per source bin 
(one for errors in the photo-$z$ distribution
and one for biases in the shear calibration) and one nuisance
parameter per lens bin (the linear bias, $b_1$; the non-linear bias, $b_2$,
accounting for scale dependence and stochasticity, is studied in Section~\ref{sec:scale}), 
plus an additional term, $\alpha$, to account for potential systematic errors 
induced by observational effects that might induce an overall shift in 
the normalisation of the amplitude of $\wtheta$ (see Section~\ref{sec:lss_systematics}).  
The full set of nuisance parameters and their priors are listed in the lower
half of Table~\ref{tab:parameters} and summarized below.

%We marginalize over two nuisance parameters per lens bin corresponding
%to the linear and non-linear galaxy bias, previosuly defined in
%Section \ref{sec:nonlinearbias}. The introduction of these two
%parameters accounts for effects such as scale dependency and
%stochasticity and we further test the impact of these in the
%results in the next section.
%JK: I don't this bit is necessary, already covered above.

\begin{itemize}
\item \textbf{Photometric redshift calibration:} For each source bin $i$, we marginalize over a
  photo-$z$ bias parameter, $\beta_i$, defined
  such as $n_i(z) \rightarrow n_i^{}\left(z+\beta_i\right)$.
%where $\beta_i$ is a free parameter to be marginalized over during the fitting process.
In \citet{bonnett15}, it was found that a single additive parameter
for the photo-$z$ distribution with a Gaussian prior centered on zero
with a dispersion of 0.05, was sufficient to account for any statistical
bias on $\Sigma_{\rm crit}$ and hence $\sigma_8$ within the degree of
statistical error expected for the SV catalogs.

\item \textbf{Shear calibration:} For each source bin $i$, we marginalize over an extra
  nuisance parameter $m_i$, to account for the shear
  calibration uncertainties, such that $\gamma_{t;i}(\theta)
  \rightarrow \left(1+m_i\right)\gamma_{t;i}^{}(\theta)$, with a
  Gaussian prior with mean 0 and width 0.05, as advocated in
  \citet{jarvis15}.
 
\item \textbf{Additive $w(\theta)$ constant:} We marginalize over an
  additive constant parameter, $\alpha$, in the galaxy angular
  correlation function: $w(\theta) \rightarrow
  w(\theta)+10^{\alpha}$. This parameter accounts for possible
  systematics arising from variations in observing conditions 
  across the field, stellar contamination and masking~\citep{ross11}, 
  which we also test for in the next section.
\end{itemize}

\begin{table*}
   \centering
   %\topcaption{Table captions are better up top} % requires the topcapt package
   \begin{tabular}{@{} lcccccccc@{}}% Column formatting, @{} suppresses leading/trailing space
      \toprule
      Probes            & $z$                                                & $100\beta_1 $          & $100\beta_2$   & $100m_1 $  & $100m_2$  & $\alpha$\\
      \midrule

      DES   ($\Lambda$CDM)                & 0.2 $<z<$ 0.35 &  $-0.89 \pm   4.58 $& $0.25 \pm 4.56    $
      	                                                                               &  $-0.09 \pm   4.59 $& $0.44  \pm 4.42$    
	                                                                               &  -3.41 $\pm$  0.84  \\
	                                                                               
      DES   ($w$CDM)                           & 0.2 $<z<$ 0.35 &  $-1.00  \pm  4.53 $& $0.13 \pm  4.51 $
                               						            &  $-0.85 \pm   4.47 $& $0.14  \pm  4.57 $ 
						                                     &  -3.42 $\pm$ 0.83\\  
                                                                                        
      DES   ($\Lambda$CDM)                 & 0.35 $<z<$ 0.5 &  $-1.77  \pm  4.46  $& $0.14  \pm  4.67  $
                               						            &  $-0.05  \pm   4.65  $& $0.36  \pm  4.64  $
						                                     & -3.57 $\pm$ 0.81\\   
						                                     
       DES   ($w$CDM)                           & 0.35 $<z<$ 0.5 &  $-1.78   \pm  4.38  $& $0.18  \pm  4.48  $
                               						            &  $-0.85    \pm  4.48  $& $0.05  \pm  4.31  $
						                                     & -3.49 $\pm$ 0.81\\   						                                     
						            
      DES + Planck ($\Lambda$CDM)   & 0.2 $<z<$ 0.35 &  $-0.58 \pm  4.83  $& $0.29 \pm  4.99  $
                               						            &  $-0.63  \pm  4.87  $& $0.72\pm  4.84  $ 
						                                     &   -3.62 $\pm$ 0.82\\    
						            
      DES + Planck ($w$CDM)               & 0.2 $<z<$ 0.35 &  $-0.87 \pm  4.73  $& $0.14\pm  4.87  $
                               						            &  $-0.76  \pm  4.88  $& $0.41 \pm  4.79  $ 
						                                     &   -3.62 $\pm$ 0.82\\  
						            
      DES + Planck  ($\Lambda$CDM)  & 0.35 $<z<$ 0.5 &  $-3.11 \pm  4.48  $& $-0.53   \pm  4.95  $
                               						            &  $-0.99\pm  4.92  $& $-0.65   \pm   4.77  $  
						                                     & -3.44 $\pm$ 0.87\\
						             
      DES + Planck  ($w$CDM)              & 0.35 $<z<$ 0.5 &  $-1.04  \pm  2.53  $& $-0.16   \pm  2.64  $
                               						             &  $-1.09\pm   4.32  $& $-0.68   \pm  4.34  $  
						                                      & -3.43 $\pm$ 0.85\\

      \bottomrule
     
   \end{tabular}

\caption{
Marginalized mean systematic uncertainty parameters with 1-$\sigma$ errors measured from the posterior distribution 
of the joint analysis of angular clustering and galaxy-galaxy lensing in DES-SV data.
   We assume a Gaussian prior (centered on zero) for each systematic parameter, while the width of
   the prior is set from~\citet{jarvis15} for the shear calibration and~\citet{bonnett15} for the photo$-z$s. 
   Each nuisance parameter is additionally truncated by the amounts in Table~\ref{tab:parameters}.}

\label{tab:systematics} 
\end{table*}

The resulting constraints in the $\Omega_m$ and $\sigma_8$ plane are
shown in Fig.~\ref{fig:comparison_shear_2pt}.  The 2D contours are
centered around $\Omega_m \sim 0.3$ and $\sigma_8 \sim 0.75$, and
marginalizing out the other parameter we find the following 1D constraints:
$\Omega_m = 0.31 \pm 0.10$ and $\sigma_8 = 0.74 \pm 0.13$.  Comparing to
measurements from Planck~\citep{planck15} and DES Cosmic Shear \citep{DES15}
alone, we are consistent at the $\sim 1\sigma$ or better level. We combine results from  
the two experiments in 
Section~\ref{sec:discussion}. In addition, we see the same direction of 
degeneracy between these two parameters as with cosmic shear, although 
the degeneracy is not quite as strong with $w(\theta)$ and $\gamt$.

We also include $w$, the dark energy equation of state parameter, 
as an additional free parameter in Fig.~\ref{fig:wCDM}. We found that the 
DES-SV data alone was unable to provide strong constraints on $w$ 
and obtained $w = -1.93 \pm 1.16$. 
%One reason for this, is that 
%this combination of measurements is more sensitive to $\Omega_m$ and $\sigma_8$
%than $w$ and the covariance between these parameters and $w$ is fairly weak. 
However, compared to Planck (red contours), the DES-SV constraints on 
$\Omega_m$ and $\sigma_8$ are degraded far less when $w$ is introduced 
as a free parameter. Also, we note that the preference for $w < -1$ values is 
determined by our choice of prior on $w$; we require $-5 < w < -0.33$, so the prior volume 
covered by $w < -1$ is greater than $w > -1$ and in the absence of a strong
constraint on $w$, values of $w < -1$ are favored.

Table~\ref{tab:results} contains a more detailed summary of our
findings for this fiducial setup, assuming either a $\Lambda$CDM or
$w$CDM cosmology.  In addition to DES $w(\theta)$ and $\gamt$, we show
results combined with Planck.  Table~\ref{tab:results} also
shows results for our lower redshift lens bin, $0.2 < z < 0.35$.  For these
results we vary only the cosmological parameters
$\{\Omega_m, \Omega_b, h, n_s, \sigma_8\}$ and $w$ where noted (in
addition to the nuisance parameters described in the present and following
sections). When combined with constraints from Planck, we also allow
the optical depth, $\tau$, to vary as well, since the CMB has additional
sensitivity to physics that is only weakly captured by large scale
clustering at late times. Table~\ref{tab:systematics} shows the constraints on 
the nuisance parameters related to photo-$z$ and shear calibration described above.

%JK: We assume a small non-zero neutrino mass
%with a fixed value of $\Sigma m_\nu = 0.06$ eV under the normal
%hierarchy for consistency with the Planck analysis.

%We marginalize over wide flat priors on several cosmological
%parameters \cs{($h$, $\Omega_b$, $n_s$? neutrino mass fixed? prior
%  ranges?)}, so we vary \cs{N} cosmological parameters assuming a flat
%universe. The results do not change substantially under the
%marginalization over these paremeters \cs{?}. Priors are chosen to be
%safely wider than existing Planck 2015 chains, with which we will
%combine our results in Section \cs{X}.
%JK: I took this out because I thought it was unnecessary. The tables and
%figures should say that the other parameters have been marginalized. 

In the following section, we will study the robustness of these
results under changes in the configuration of the data vector and the
systematics modelling.

%%%%%%%%%%%%%%%%%%
\begin{figure*} %  figure placement: here, top, bottom, or page
   \includegraphics[width=\linewidth]{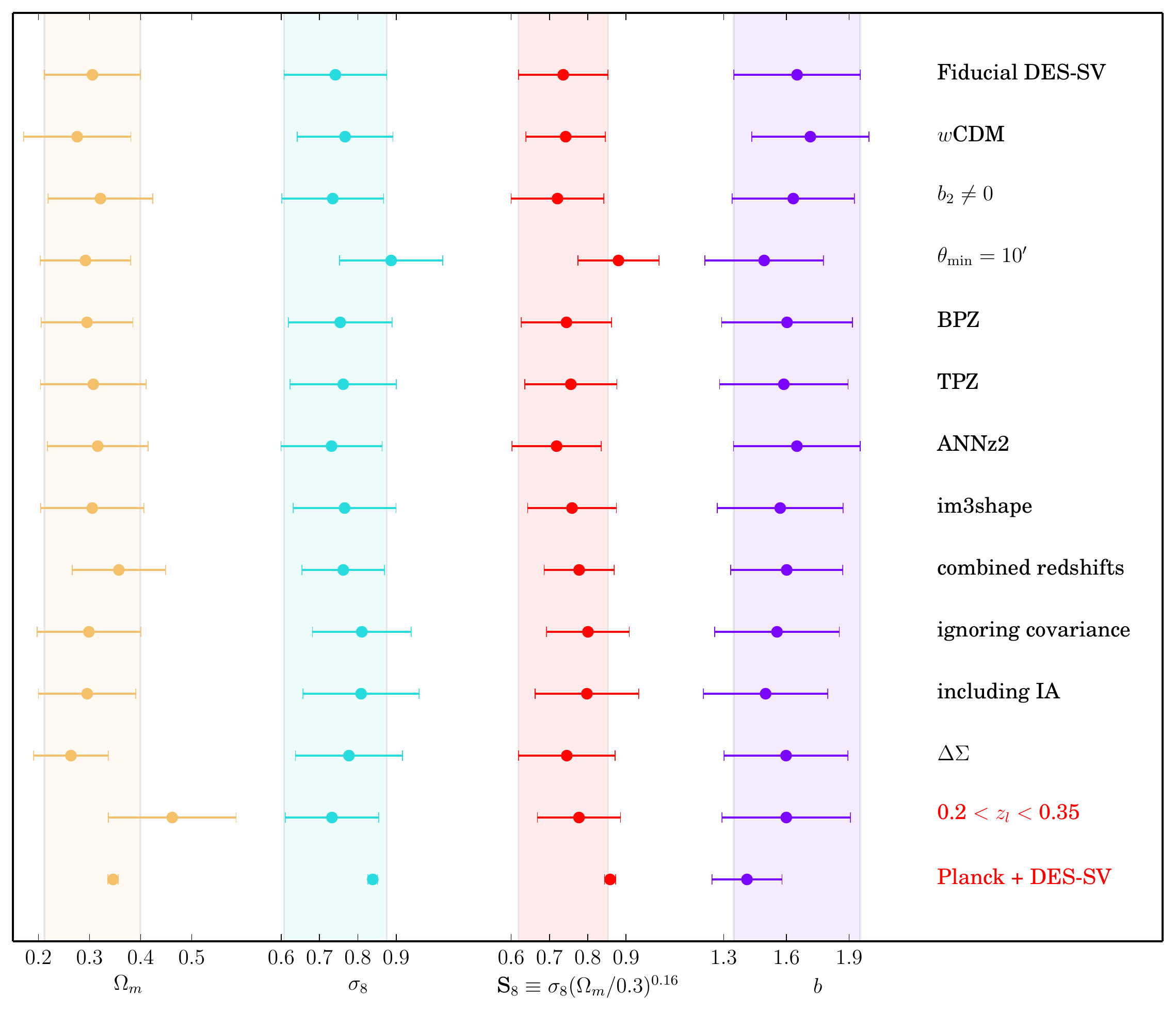} 
   \caption{Marginalized 1D posterior constraints on $\{\Omega_m,
     \sigma_8, S8, b_1\}$ for the lens bin 0.35 $< z < $ 0.5 for various
     configurations in our pipeline. For this figure, we have defined S$_8 \equiv \sigma_8(\Omega_m/0.3)^{0.16}$, 
     that is, we hold the index fixed to the degeneracy direction found for our fiducial analysis. Note that 
     this value is substantially different to one favoured by Planck data alone, but we have chosen a
     constant value to enable comparisons between the systematic tests. 
     Our fiducial results use shear catalogs from \texttt{ngmix}, SkyNet photometric redshifts, and linear bias
     in a $\Lambda$CDM cosmology, as described in Section~\ref{sec:cosmology}. 
     The different rows in this plot are obtained by 
     varying the fiducial assumptions individually to test their
     impact on the parameter constraints, and they are all detailed in Section~\ref{sec:testing} and the Appendix.
     Tests involving (nearly) independent data are highlighted in red near the end of the table.}
   \label{fig:constraints_table}
\end{figure*}

%%%%%%%%%%%%%%%%%%%%
\section{Robustness of the results}\label{sec:testing}

%Having presented the fiducial cosmological results in this work, we
%must verify that our measurements of angular clustering and tangential
%shear and cosmological analyses are robust against systematic
%errors. 
In this section, we describe the suite of tests performed to
check that our conclusions are unbiased with respect to errors in the
shear and photo-$z$ calibrations, intrinsic alignments, survey geometry, 
choice of angular scales and theoretical modelling of the data vectors. 
The results in this section are displayed in Fig.~\ref{fig:constraints_table},
for the parameters we are most sensitive to in this work: $\{\Omega_m, \sigma_8, b_1\}$.
The different rows correspond to the different tests described
in this section or in the Appendix, where we check the results from a different lensing estimator.
Despite the changes in the photo-$z$ algorithms, the shear catalogs, 
the weighting of the lens-source pairs, non-linear bias modelling and 
choice of scale, our estimates for these cosmological parameters in 
Fig.~\ref{fig:constraints_table} usually remain within 1-$\sigma$ 
of the fiducial constraints.

A number of systematics that are unique to the measurement of the
tangential shear such as the calibration of galaxy ellipticities, the
effect of different shear calibration pipelines, null detection of the
lensing B mode and effect of photo-$z$ errors in the lens and source
catalogs on the measurement have already been accounted for in~\citet{clampitt16}, 
so we do not present tests for these effects again.
For more information on tests of the shear pipeline, we
refer the reader to \citet{jarvis15} while \citet{bonnett15}
contains extensive tests of the photo-$z$ calibration algorithms. We
also check for possible systematics introduced by the effects of survey 
geometry, depth and varying observing conditions in the survey following the
techniques in \citet{crocce15}. 

Our analysis pipeline accounts for the effect of a number of
systematics which are folded into our final constraints on cosmology.
To first order, these nuisance parameters are responsible for 
altering the amplitude of $\wtheta$ and $\gamt$, and so are 
strongly degenerate with one another. As a result, we were unable to
constrain these parameters beyond their prior distributions and the results
in Table~\ref{tab:systematics} show that the posterior distributions of the 
nuisance parameters no more informative than the priors.
To determine which of these most affect our results, we have analysed
each of these systematics individually by running chains in four
scenarios: no systematics, shear calibration only, photo-$z$ errors only,
full weak lensing systematics but no constant offset in w($\theta$),
and shear calibration with photo-$z$ errors (our fiducial set up).  We 
found that including an additive constant to $w(\theta)$ was responsible 
for the greatest decrease in precision on the 1D marginalized 
constraints on $\Omega_m$, with the 1-$\sigma$ error on $\Omega_m$
increasing by as much as 17\% compared to the no systematics case.
However,  $\sigma_8$ was much less affected with a difference below 3\%. 
In comparison, accounting for photo-$z$ errors with an additional two
free parameters in the $N(z)$ distribution increased the error on both
parameters by about 8\%. The change from including two shear 
calibration parameters was smaller still, with only a 3\% reduction
in precision for $\Omega_m$ and 5\% for $\sigma_8$ relative to the 
no systematics case. We also found small changes to the best 
fitting values, well within the 1-$\sigma$ confidence interval, 
as expected from Fig.~\ref{fig:constraints_table}.

\subsection{Choice of scales}
\label{sec:scale}
There are several reasons to limit the range of scales that we consider in our analysis. 
The large scale cutoff is set by the size of the SV patch and how well the geometry 
of the region can be modelled; we found that our jackknife estimates of the 
covariance matrix overestimated the covariance matrix obtained from 
50 independent N-body simulations above $70'$ (see Fig 5, \citealt{clampitt16}).

On small scales, we are limited by how well we can model the nonlinear
clustering of matter and of \redmagic{} galaxies. Galaxy formation preferentially
occurs in high density environments within dark matter halos and is
subject to a number of complex baryonic processes; 
these are not captured in our model predictions for the mass power spectrum and 
potentially introduce a non-trivial bias between the dark matter
and the galaxies. This is particularly important for the tangential shear, 
which contains a mixture of small and large scale information; i.e. imposing
a sharp cutoff in angular scale does not completely eliminate the effect
of scales below that cutoff~\citep{mandelbaum13}. On small enough 
scales, we expect to observe effects such as stochasticity, non-local bias 
and scale dependence. These could invalidate the linear bias model 
used in our analysis. 
%assumption of linearity in Eqn.~\ref{eqn:pow}
%that allows us to relate the clustering of dark matter to galaxies.

%There are numerous approaches to populating the dark matter
%distribution with galaxies, such as using Halo Occupation Distribution
%(HOD) modelling, abundance matching, semi-analytic galaxy formation
%models, and perturbation theory \jk{[refs]}. We would like to avoid
%using a HOD approach which can introduce as many as six additional
%free parameters, while abundance matching and semi-analytics really
%only work in concert with a N-body simulation in which the dark matter
%halo properties are known. Thus, we have adopted the perturbation
%theory scheme presented in Section~\ref{sec:theory} as our approach.
%In this model, we have an additional two free parameters, $b2$ and $N$, 
%which represent the scale dependence of the biasing and the shot noise
%respectively. 

\begin{figure} 
\centering
   \includegraphics[width=\linewidth]{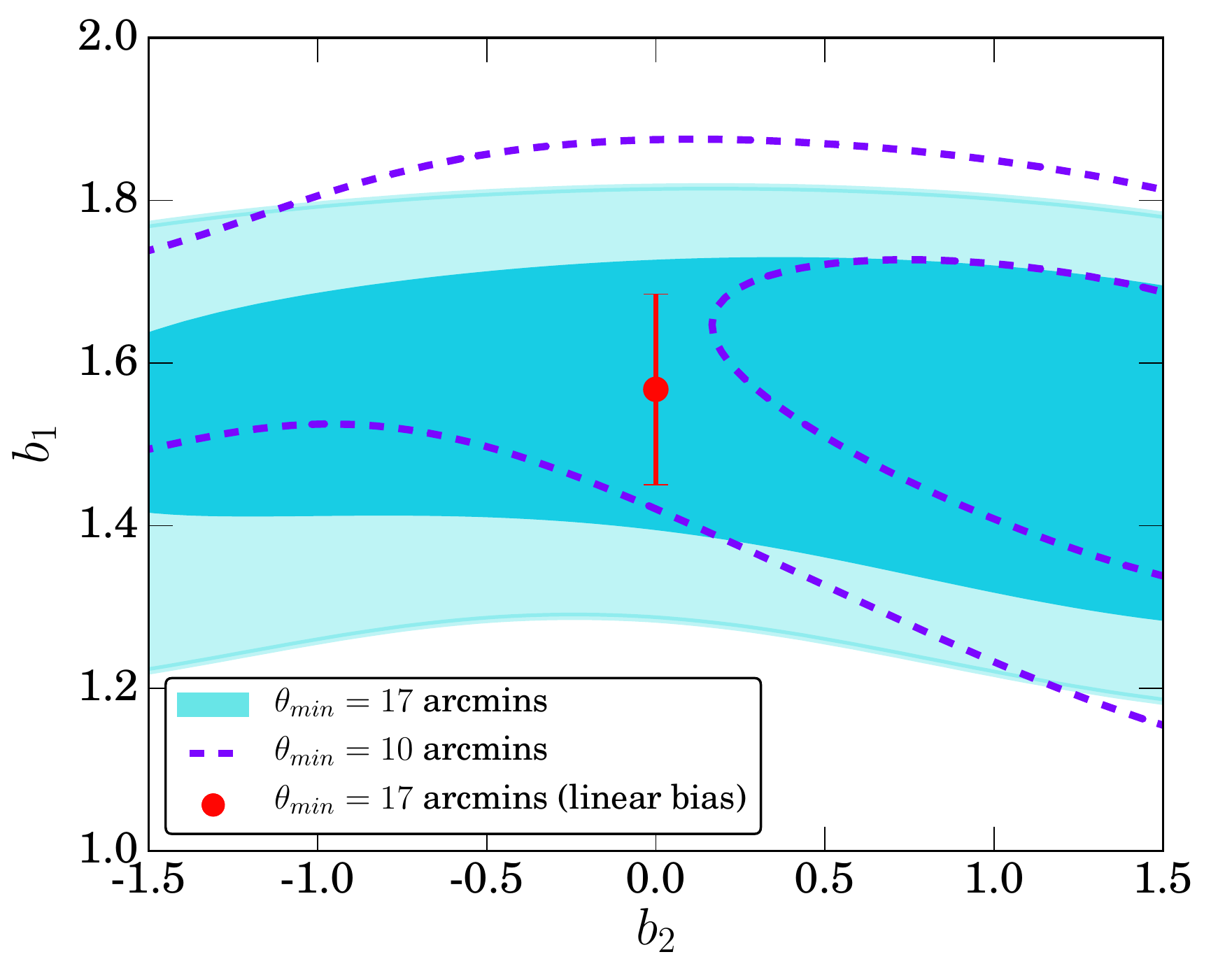} 
   %\includegraphics[width=\linewidth]{b1_b2_sims_test.pdf} 
   %\resizebox{120mm}{!}{\includegraphics{wtheta_sims_testing.png}}
   \caption{The posterior distribution on the bias parameters, $b_1,b_2$ 
   from simulations of $w(\theta)$ for the redshift bin 0.35 $< z <$ 0.5. 
   We fit the~\citet{mcdonald06} model to a minimum cut in scale at 
   $10'$ (cyan) and $17'$ (purple) and a linear bias model to $17'$ (red point)
   to demonstrate the insensitivity of our fiducial results with a $17'$ cutoff 
   to $b_2$. }
   \label{fig:wtheta_sims_testing}
\end{figure}

In this section, we present simulation based tests to determine the
smallest scales for which the linear bias model and perturbation 
theory model of~\citet{mcdonald06} are valid. We use a mock catalog 
designed to reproduce the properties of the DES-SV survey.
The catalog is based on an N-body simulation (c-400; see also \citet{mao15, 
lehmann15}) run with the \textsc{L-Gadget} code, a variant of \textsc{Gadget}~\citep{springel05}.
The simulation has a box size of 400 Mpc$/h$ with 2048$^3$ particles and 
a force resolution of 5.5 kpc$/h$.  Halo catalogs were generated with the
\textsc{Rockstar} halo finder \citep{behroozi13a} and the \textsc{Consistent Trees} 
merger tree builder~\citep{behroozi13b}. A galaxy catalog was produced using 
an abundance matching technique, as described in~\citet{reddick13} and 
\citet{lehmann15}, with halos ranked according to the peak halo velocity and 
assigned a luminosity from the \citet{blanton03} luminosity function, using a scatter of 0.2 dex.
Snapshots from the simulation were combined into a lightcone with the same footprint as the
DES-SV region.  Galaxy colors were assigned using the empirically derived relationship between
luminosity, projected distance to the fifth nearest neighbor galaxy, and galaxy SED (this method
for assigning colors has been used in previous generations of catalogs, see e.g. \citealt{cunha12, chang15}).
Photometric errors were added to match the depth distribution of DES-SV galaxies.  
The \redmagic{} algorithm was run on the lightcone, using the same technique as applied to the 
DES-SV data and this produced a mock \redmagic{} catalog. The \redmagic{} color model 
is retuned to the simulations before identifying these galaxies, but was found to 
have similar properties to that seen in the data. We find that the clustering properties 
of the \redmagic{} galaxies in this catalog are consistent with those measured in DES-SV data.

From the mock catalog, we have measured $w(\theta)$ in the same bins in
redshift, 0.2 $< z <$ 0.35 and 0.35 $< z <$ 0.5, from $10' < \theta <
100'$. Our covariance matrix is calculated from a jackknife resampling
of the catalog as described in Section~\ref{sec:covariances}.

We test our bias modelling by making two cuts in angular scale at $10'$ 
and $17'$, corresponding to $(\sim 3$ Mpc$/h)$ and $(\sim 5.5$ Mpc$/h)$, 
because we expect the bias to transition between its large scale asymptotic
limit to scale dependence somewhere in this regime for the galaxy type
that we consider. We fit both a linear and a quasilinear bias model with 
two free parameters, $b_1$ and $b_2$, as described in Section~\ref{sec:nonlinearbias}
to the simulated $w(\theta)$ while holding the cosmological parameters
fixed to the value of the N-body simulation. 
Note that the effect of the shot noise parameter, $N$, on $w(\theta)$ is 
negligible on our scales of interest so we do not include it in our tests.
Figure~\ref{fig:wtheta_sims_testing} shows the recovered biases when all 
the cosmological parameters are fixed at the simulation values
for the fiducial lens bin of 0.35 $< z <$ 0.5. The measured $w(\theta)$
is insensitive to the value of $b_2$ when a minimum angular scale 
of $17'$ is chosen (cyan filled contour) and we are simply recovering our 
prior distribution on $b_2$. 

When we change the minimum scale to $10'$ (purple dashed contour), 
there is a 1-$\sigma$ preference for a non-zero value. Using a linear model 
of biasing (Fig.~\ref{fig:wtheta_sims_testing}; red point) with the same 
fixed cosmology set up, we find that we recover the same value 
of $b_1$ as in the non-linear case. We obtain similar results for the low-$z$ lens bin, 
except that the minimum scale cutoff is now at $22'$ for $w(\theta)$ to be well 
modelled by a linear bias. Figure~\ref{fig:wtheta_sims_testing}  demonstrates 
that our choice of using a linear bias up to these angular scales for the \redmagic{} 
sample should not affect our ability to constrain cosmology. Based on 
these results, we can conclude that applying a linear bias model with
$\theta_{\rm min} = 17'$ ($22'$) for the high$-z$ (low-$z$) lens bin will 
not bias our results in the presence of scale dependent non-linear biasing. 
As an additional check, we have also rerun our fiducial analysis with $b_2$ as
an additional free parameter. For these fits, we obtained $\Omega_m$ = 
0.32 $\pm$ 0.10,  $\sigma_8$ = 0.73 $\pm$ 0.13, $b_1 = 1.63 \pm 0.29$ 
and $b_2 = -0.14 \pm 0.76$, which is consistent with our fiducial results.

For the shear catalogues, \citet{jarvis15} identified $3'$ as the 
angular scale in the shear auto correlation function at which the 
additive errors contribute to half of the total 
forecasted error on the measurement of $\sigma_8$ or about $\sim 3$\%. 
Although it is expected that position-shear correlations are less sensitive to
 additive systematics in the shear, 
 we only consider angular scales 
 $\theta \geq 10'$ even for the tests of the bias model above. 
This $3'$ cutoff is well outside of the minimum scales considered in 
our cosmological analysis which use at most $\theta>17'$.

%%%%%%%%%%%%
\subsection{Photo-$z$ systematics}

Since DES-SV is an imaging survey, the quality of our constraints rely
heavily on being able to robustly calibrate the photometric redshifts
of the lens and source galaxy samples. However, because $\wtheta$ does
not use radial information, apart from the selection function, it is
relatively insulated from photometric errors compared to the full
3D correlation function. Furthermore, because the photometric error in the lens
\redmagic{} sample is so small~\citep{rozo15b}, the potential
systematic errors in the cosmology analysis are dominated by the
photometric redshifts of the source galaxy sample.

\begin{figure} 
   \centering
   \includegraphics[width=\linewidth]{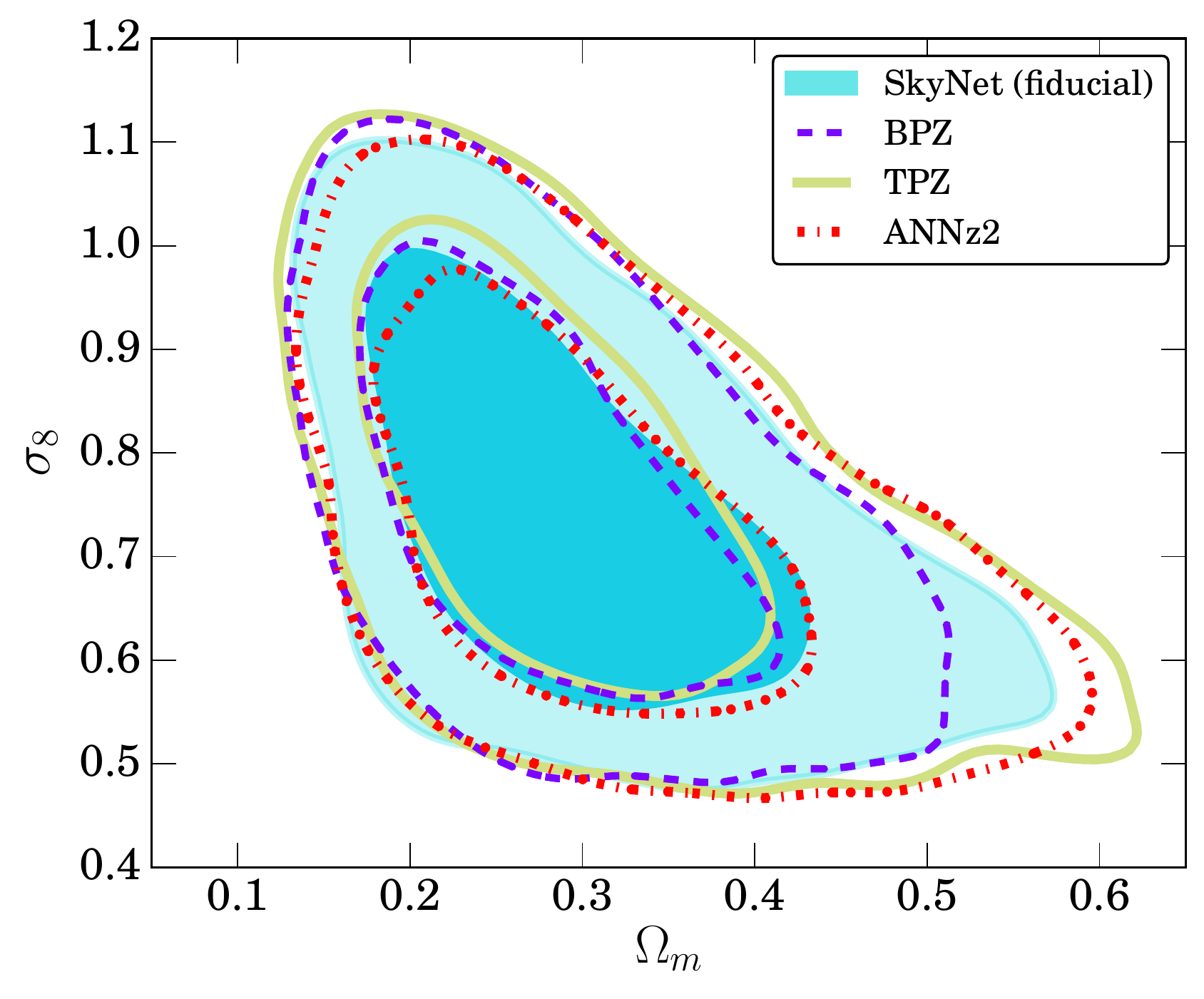} 
   \caption{Constraints on $\Omega_m$ and $\sigma_8$ using four
     different photo-$z$ codes to calculate the redshift distribution of
     sources. The contours for the 0.35 $< z <$ 0.5 redshift bin are shown
     here; we have also  checked the robustness of our
     results for lenses in the redshift bin 0.2 $< z <$ 0.35. 
       }
   \label{fig:photo_z_test}
\end{figure}

We deal with photometric redshift systematics in two different ways. 
First, we follow the recommendations of \citet{bonnett15} and define 
an additional photo-$z$ bias parameter for each source bin, $i$, as:
\begin{equation}
n_i^{pred}(z) = n_i^{obs}\left(z+\beta_i\right) 
\label{eqn:photoz}
\end{equation}
where $\beta_i$ is a free parameter with a Gaussian prior of width
0.05 to be constrained during the fitting process. The width of the
prior is set to be consistent with~\citet{bonnett15}, where it was found
that the difference between photometric and spectroscopic estimates of
the redshift of the training samples that most closely resemble our
shear catalogs have a relative mean bias with a Gaussian dispersion
of 0.05. This method was also used in the DES-SV Cosmic Shear 
Cosmology paper~\citep{DES15}. We found that introducing an additional 
photo-$z$ bias parameter for each source bin increases our uncertainty 
by, at most, 8\% compared to the constraints we would have if we did 
not fit for any systematic parameters.

In addition, we check that our constraints are robust to our choice of
photo-$z$ algorithm. Our fiducial shear catalogs use photometric
redshifts derived from the SkyNet algorithm \citep{graff14,bonnett15b}, and we
have repeated our analysis by using the redshift distribution given by
three other photo-$z$ codes studied in \citet{bonnett15}, namely BPZ,
TPZ and ANNz2.  For this test, we assume a $\Lambda$CDM cosmology and allow
the cosmological parameters $\{\Omega_m, \Omega_b,
h, \sigma_8, n_s, b_1\}$ to vary.  In addition, we also fit for the usual systematic
parameters, $\beta_i$ for the photo-$z$ bias and $m_i$ for the
multiplicative bias in the shear calibration and the same prior
distributions. The resulting constraints in Fig.~\ref{fig:photo_z_test} 
(and \ref{fig:constraints_table}) show that our results are insensitive 
to the choice of the photo-$z$ algorithm.

Interested readers should refer to \citet{bonnett15} for a full discussion of
the photo-$z$ methods considered and the systematics modelling that we
have only summarized here.

%%%%%%%%%%%%%%%%
\subsection{Shear calibration systematics}
\label{sec:shear_sys}

Here we present our approach to modelling a possible residual
error in the shear calibration. For the interested reader, the full
details of the production and testing of the shear catalogs used in
this analysis are given in \citet{jarvis15}.

%In [Jarvis], it was reported that the multiplicative bias is less than 0.03 for angular scales ??? for shear-shear/position-shear correlations [check details]. This sets the minimum scale that we can reasonably use for our analysis. 
Similar to the photometric redshift case, we deal with potential shear
calibration systematics on two fronts.  Firstly, we include an extra
nuisance parameter for the shear calibration, $m_i$, as:
\begin{equation}
\gamma_{t;i}^{pred}(\theta) = \left(1+m_i\right)\gamma_{t;i}^{obs}(\theta)
\label{eqn:shear}
\end{equation}
with a Gaussian prior, $p(m_i)$, with mean 0 and width 0.05, for each source bin $i$ in our
analysis as recommended in \citet{jarvis15}. Contamination from additive
errors in the shear estimation are expected to be minimal for
galaxy-galaxy lensing, because of the azimuthal symmetry of the lens
system. Including an additional parameter for the shear calibration 
degrades our constraints by, at most, 5\%, compared to all systematic
parameters being ignored or set to fixed values. 

Secondly, galaxy images in the DES-SV region were analyzed with two
pipelines, {\ngmix} and \texttt{im3shape}.  \citet{jarvis15} showed
that they both produced consistent results that satisfied the SV
requirements for weak lensing, i.e. that less than half of the
forecasted error on $\sigma_8$ (about 3\%) originates from systematics
in the measurement of the shear.  Although we have chosen to use the
\texttt{ngmix} catalog for our analysis, we have also rerun the
analysis pipeline on the \texttt{im3shape} catalog to check that our
results are not sensitive to the shear catalog used (see
Fig.~\ref{fig:ngmix_im3shape} for a comparison of lensing measurements
using the two shear pipelines). We found that the cosmological 
parameters varied imperceptibly when the \texttt{im3shape} catalog 
was used instead of \texttt{ngmix}. This is demonstrated in 
Fig.~\ref{fig:constraints_table}.

\subsection{Intrinsic Alignments}
Correlations between the intrinsic shapes and orientations of lensing
sources, known as ``intrinsic alignments'' (IA), are one of the most
significant astrophysical sources of uncertainty in weak lensing
measurements (see \citealt{troxel15,joachimi15} for recent
reviews). Although typically considered in the context of shear-shear
correlations, IA can also contaminate galaxy-galaxy lensing
measurements due to uncertainties in photo-$z$ estimates which lead to
overlap in the true lens and source distributions (see Fig.~\ref{fig:nzs}). The intrinsic
shapes of sources can be correlated with the positions of lenses at
the same redshift \citep{blazek12}.

In general, the contamination from IA reflects the (potentially
nonlinear) relationship between large-scale structure and galaxy
shapes, as well as the clustering of lenses and physically associated
sources. However, observational evidence
(e.g.\ \citealt{joachimi11,blazek11,singh15}) indicates that the
dominant IA contribution is likely from elliptical
(pressure-supported) galaxies, for which the IA component is linearly
related to the large-scale tidal field. This ``tidal alignment''
paradigm \citep{catelan01,hirata04,blazek15} was recently used to
mitigate IA in the DES-SV Cosmic Shear Cosmology analysis \citep{DES15}. In this
work, we consider scales on which the clustering of lens-source pairs
is negligible (see \citealt{clampitt16} for further discussion). In
this regime, tidal alignment predicts that the fractional IA
contamination to the lensing signal is nearly scale-invariant. Both
the IA and lensing are sourced by the same matter power spectrum, even
in the presence of nonlinear evolution, and we find that the different
line-of-sight weighting for IA and lensing (e.g.\ Eq.~\ref{eqn:tangential_shear})
leads to negligible relative scale-dependence in angular correlations.

We thus account for the potential impact of IA in our 
analysis by including an additional term that modifies the amplitude
of the tangential shear, such that $\gamt \rightarrow (1+m_{\rm shear
  \; cal}+m_{\rm IA})\gamt$. We place a Gaussian prior on $m_{\rm IA}$ of 8\%
$\pm 4$\% for the lower redshift source bin. The higher redshift source bin is
  sufficiently separated from the redshift of the lenses that the potential IA contamination is
  negligible. Our priors on the expected IA contamination are
  calculated from the overlap in lens and source redshift
  distributions and assume an IA amplitude consistent with the
  constraints found in the cosmic shear analysis of the same sources
  on the DES-SV patch \citep{DES15}. Potential IA contamination in the
  galaxy-galaxy lensing measurement is discussed further
  in~\citet{clampitt16}.

We do not observe a significant detection of IA contamination beyond the
prior imposed;  we find that $m_{\rm shear\; cal, 1}+m_{\rm IA, 1} \sim8.0 
\pm 3.7$\% for the low redshift sources with $m_{\rm shear\; cal, 2} \sim -5.3 \times 10^{-4} \pm
4.5$\% for the higher source bin. Including IA only affects the cosmology results
by, at most, inducing a $\sim$3\% shift towards a lower value of $\Omega_m$ 
compared to the fiducial case without IA, as shown in Fig~\ref{fig:constraints_table}. 
For $\sigma_8$, the change was much smaller, with a fractional shift of less than a
percent. Because the inclusion of IA contamination has a negligible effect on 
our results, compared to the statistical errors, we do not include IA modelling 
for our fiducial analysis. 

%%%%%%%%%%%%%%%%%
\subsection{Impact of observing conditions}
\label{sec:lss_systematics}

\begin{figure}
\centering
\resizebox{90mm}{!}{\includegraphics{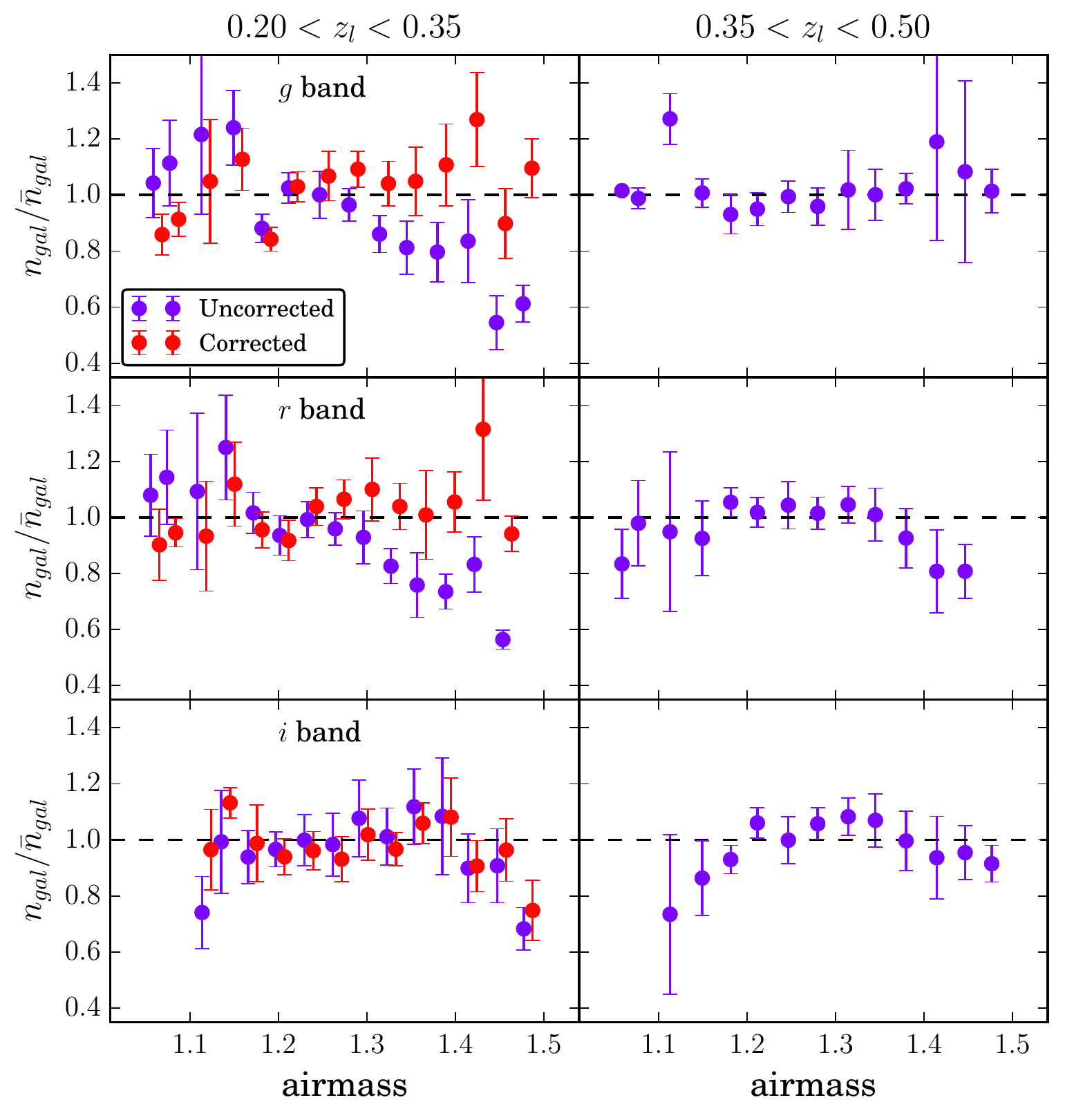}}
\caption{\redmagic{} galaxy density as a function of \texttt{airmass}
  in $g$, $r$ and $i$ bands for the two lens redshift bins considered
  in this work. A significant correlation is present for the $g$ and
  $r$ bands in the low-$z$ bin, which we correct by
  weighting galaxies inversely by the \texttt{airmass}
  values at the sky position. Note that we do not apply a correction to the 
  high-$z$ bin since it doesn't show a significant correlation with any 
  systematics parameter.}
\label{fig:lss_sys}
\end{figure}

Photometric galaxy surveys such as DES are affected by time-dependent
fluctuations in observing conditions that may impact the galaxy
catalogs. There are a number of effects that can modulate the
detection efficiency of galaxies and cause density variations across
the survey footprint.  In this section we follow the approach of
\citet{crocce15} and consider single-epoch properties that affect the
sensitivity of the survey and hence may affect the galaxy clustering and
galaxy-galaxy lensing observables. We use projected
HEALPix\footnote{\texttt{http://healpix.sf.net}} \citep{gorski05} sky
maps (with resolution \texttt{nside}=4096) in $grizY$ bands for the
following quantities:
\begin{itemize}
\item \texttt{depth}: mean survey depth, computed as the mean magnitude for which galaxies are detected at $S/N=10$. 
\item \texttt{FWHM}: mean seeing, in pixel units, computed as the full width at half maximum of the flux profile. 
\item \texttt{airmass}: mean airmass, computed as the optical path length for light from a celestial object through Earth's atmosphere (in the secant approximation), relative to that at the zenith for the altitude of CTIO.
\item \texttt{skysigma}: mean sky background noise, computed as the flux variance per amplifier in chip of the CCD. 
\item \texttt{USNO}: mean stellar density, as measured by the USNO-B1 stellar catalog \citep{monet2003} with B magnitude brighter than 20 to ensure constant depth across the field. 
\end{itemize}
See \citet{leistedt15} for a full description of these maps.

We study the density of \redmagic{} galaxies in the two lens bins as a function 
of each of these quantities that can potentially result in systematic effects. 
To ensure the data is free of such systematics, we require the galaxy density 
to be uncorrelated with the observed \texttt{depth}, \texttt{FWHM}, \texttt{airmass}, 
 \texttt{skysigma} and \texttt{USNO}, otherwise we apply a correction to remove the dependency.  
Among the five quantities for each band and each lens bin considered here, 
we only find a significant correlation in the low-$z$ bin with \texttt{airmass} 
in the $g$ and $r$ DES bands. This trend is demonstrated in Fig.~\ref{fig:lss_sys}, 
which shows the \redmagic{} galaxy density as a function of \texttt{airmass} in 
$g$, $r$ and $i$ bands for the two lens bins. In order to correct for this correlation,   
we weight galaxies according to the inverse of a linear fit to the observed trend of 
\texttt{airmass} in the $g$ band. This procedure is similar to that applied in 
\citet{ross12,ross14} to correct for systematic relationships with stellar density 
and \texttt{airmass}. The corrected results are shown in Fig.~\ref{fig:lss_sys}, 
where we see that the $g$ band weighting also corrects the trend in the $r$ band, 
as expected given the correlation present among the \texttt{airmass} maps in the 
$g$ and $r$ bands.    

In addition to the weighting correction described above, we have also applied the 
procedure used in \citet{crocce15}, in which galaxy and systematics maps are 
cross-correlated and used to correct the galaxy correlation functions. 
At the galaxy clustering level, the two approaches yield consistent results. 
Furthermore, in both cases the correction is compatible with an additive constant 
in the angular galaxy clustering signal. Nonetheless, we introduce an additive constant 
as a systematics parameter in the corrected measurement of $w(\theta)$ as outlined in 
Section~\ref{sec:cosmology} to deal with any residual systematic effects. 
This is marginalized over in the cosmological analysis according to the prior defined in 
Table \ref{tab:parameters}. On the other hand, the impact of the \texttt{airmass} correction 
in the galaxy-galaxy lensing observables is not significant given the statistical power 
of these observations in DES-SV.  
     
%Correlation at the two point level, consistency between weighting and cross-correlations. Correction compatible with an additive constant in the angular galaxy clustering signal, constant that we allow for. No effect in gglensing observables. 

As opposed to \citet{crocce15} we do not find the \texttt{depth} and \texttt{FWHM} 
maps to be relevant for our lens sample, mainly because \redmagic{} galaxies are 
much brighter than the DES main galaxy sample (Benchmark) considered in that 
work. On the other hand, correlations between \texttt{airmass} maps and galaxy positions 
were not found to be a significant systematic in \citet{crocce15}, while for \redmagic{}
galaxies in the low-$z$ lens bin, this was the only observing condition with a substantial
impact on clustering. 
While \citet{crocce15} includes all types of galaxies, the~\redmagic{} selection 
process preferentially chooses red galaxies as described in Section~\ref{sec:data}. 
It is plausible that these galaxies are more affected by \texttt{airmass}, via 
their sensitivity to atmospheric extinction. At high \texttt{airmass}, the filter bandpasses shift to the red and the RedMapper color selection, in which \redmagic{} relies, do not compensate for this. The effect is more important for the bluer DES bands $g$ and $r$~\citep{2016arXiv160100117L},
and the key spectral features of red galaxies, like the 4000\angstrom~break,  
fall in a bluer window of the filter set at lower redshifts, and hence the 
effect of atmospheric extinction is enhanced for our low-$z$ lens bin. 

In the following subsection, we present cosmology results with the low-$z$ 
lens bin after correcting for the correlation with \texttt{airmass}.

%One potential reason for this systematic is the fact that airmass
%correlates directly with atmospheric extinction, which mainly affects
%the bluer DES bands $g$ and $r$ \jc{(Li et al., in prep, des-docdb docid=8504)}
%and \redmagic{} being a photo-$z$ optimized LRG
%selection that may affect stronger the low-$z$ lens bin in which the
%4000\angstrom~break falls in a bluer window of the filter set.
%This is expected to have a minor effect in the galaxy sample of \citet{crocce15},
%since that includes all types of galaxies and does not rely on a fit
%to a calibrated red-sequence template to allow a galaxy into the sample.

\subsection{Low-$z$ lens bin results}
In this section we present the cosmology results obtained for the
low-$z$ \redmagic{} lens bin ($0.20<z<0.35$), described in Section
\ref{sec:measurements} and for which measurements are shown in
Fig.~\ref{fig:measurements}. For this bin, a significant correlation
of the galaxy density with \texttt{airmass} was found and corrected for in
Section \ref{sec:lss_systematics}. 

The photo-$z$ and shear systematics treatment in the cosmology 
pipeline is equivalent to that of the fiducial lens bin and we use 
these results as another robustness check for the cosmological 
analysis performed in this work.

\begin{figure} 
   \centering
   \includegraphics[width=\linewidth]{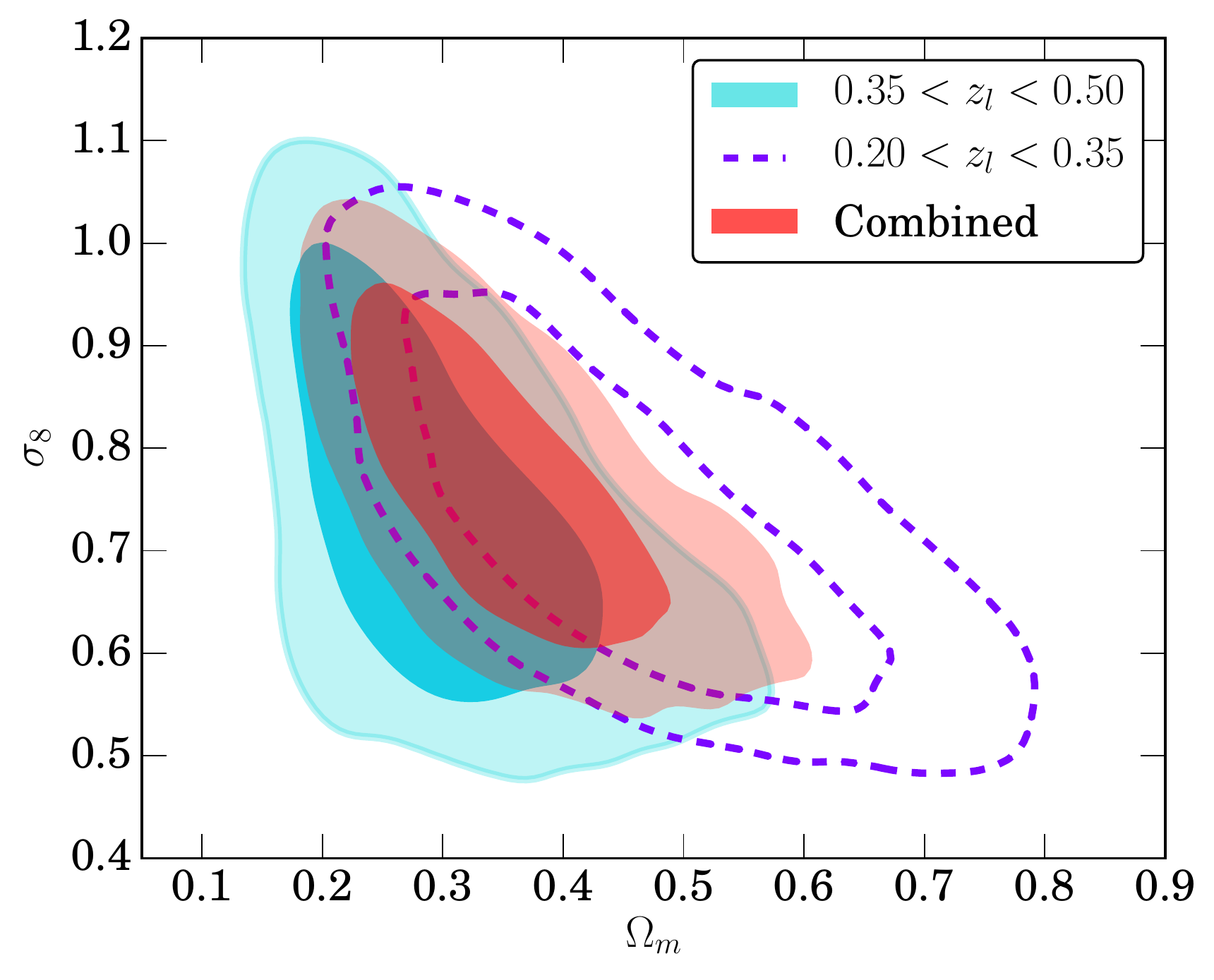} 
   \caption{Constraints on $\Omega_m$ and $\sigma_8$ using DES-SV
       $w(\theta) \; \times\; \gamt $. The fiducial high-$z$ lens bin is shown in filled blue, 
the low-$z$ lens bin is shown as dashed purple lines and the combination of the two lens bins is shown in filled red. In each case,
       a flat $\Lambda$CDM model is assumed. }
   \label{fig:lowz}
\end{figure}

The cosmological constraints obtained from these measurements are
shown in Fig.~\ref{fig:constraints_table} and Table~\ref{tab:results}, and the constraints on $\Omega_m$ and $\sigma_8$ from the combination with the fiducial high-$z$ lens bin are shown in Fig.~\ref{fig:lowz}. 
For most of the parameters, these lower redshift lenses are in agreement 
with our fiducial setup, but $\Omega_m$ shows a preference for higher 
values after correcting for the observing conditions described in 
Section~\ref{sec:lss_systematics}. Still, the results for both lens bins 
are within 1$\sigma$ of each other. 
  
Having confirmed that the results from both the low and high redshift
lens bins are consistent, we explore fitting them jointly in the same 
analysis pipeline to improve our constraints on cosmology. 
The covariance between lens bins may include a contribution from shape noise  in the shear catalog. 
We estimate
this contribution by introducing a 
random direction to the measured ellipticities before calculating the tangential
shear. This is performed $\sim$300 times to obtain a jackknife estimate 
of the shape noise across lens and source bins. We then add the shape
noise as an off-diagonal component to the covariance matrix between 
lens bins with the diagonal components being the usual JK covariance matrices
used for individual fits. We find that the marginalized 
constraints are $\Omega_m = 0.36 \pm 0.09$  and $\sigma_8 = 
0.76 \pm 0.11$, which show very little improvement on our
fiducial results. However, the constraint on S$_8 \equiv \sigma_8(\Omega_m/0.3)^\alpha$,
where $\alpha$ is chosen to be perpendicular to the degeneracy direction in the $\Omega_m$-$\sigma_8$ plane, 
shows a reduction in the error, from S$_8 = 0.735\pm0.117$ ($\alpha$ = 0.16; high-$z$ lenses only)
to S$_8 = 0.782\pm0.088$ ($\alpha = 0.21$; all lenses). 
These values of $\Omega_m$, $\sigma_8$ and S$_8$ are shown in Fig.~\ref{fig:constraints_table}. 
We do not however consider
this arrangement as our `fiducial' model, leaving joint constraints to future work with additional survey area. 

\section{Discussion}
\label{sec:discussion}
We have presented our baseline cosmological results from DES data in
Section~\ref{sec:cosmology}, assuming a flat $\Lambda$CDM model in
Figure~\ref{fig:comparison_shear_2pt} and a flat $w$CDM model in 
Figure~\ref{fig:wCDM}. 
Our results for the marginalized mean parameter values are contained in 
Table~\ref{tab:results} for each lens bin, with and without external data sets. 
%and for both $\Lambda$CDM and $w$CDM cosmologies.  
We also show 
results for each of the nuisance parameters used in our fits in 
Table~\ref{tab:systematics}. 
%We expect the values of the nuisance 
%parameters to be robust to the redshift and angular binning of the lenses 
%(the systematics in the shear catalogs should be independent of the lens selection), 
%and to within 1$\sigma$ errors, this is generally the case.

\begin{comment}
\begin{figure}[thbp] 
   \centering
   \includegraphics[width=\linewidth]{fig9.pdf} 
   %\resizebox{90mm}{!}{\includegraphics{omega_m_sigma_8_high_z_shear_2pt_planck2015_wcdm.pdf}}
   %\includegraphics[width=\linewidth]{sigma_8_omega_m_high_z_w_shear_2pt.png} 
   \caption{Constraints on $\Omega_m$ and $\sigma_8$ assuming a
       $w$CDM model using DES-SV Cosmic Shear (dashed purple), DES-SV 
       $w(\theta) \; \times\; \gamt $ (this work, blue) and Planck 2015 using only
       temperature and polarization data (TT+lowP, red). }

   \label{fig:wCDM}
\end{figure}
\end{comment}
%and these parameters only
%have a weak dependence on $w$ for these probes (see
%Figure~\ref{fig:contours}). Furthermore, the contours for $w > -1/3$
%remain unclosed; this upper bound on the equation of state is imposed
%by our prior for an accelerating universe.
%In the remainder of section, we will consider the DES data jointly
%with other datasets and also in the context of other surveys.

\subsection{External Datasets}
We performed a joint analysis of our measurements with the Planck 2015
temperature and  polarization auto and cross multipole power spectra,
$C^{TT}(\ell), C^{TE}(\ell), C^{EE}(\ell)$ and
$C^{BB}(\ell)$. Specifically, we use the full range of $C^{TT}(\ell)$
from 29 $<\ell<$ 2509 and the low-$\ell$  polarization data from 2
$<\ell<$ 29, which we denote as
Planck (TT-lowP). The inclusion of the  maps allows for
stronger constraints on $\tau$ which in turn affects $A_s$, the primordial
power spectrum amplitude. We have also chosen this configuration to allow 
for an easy comparison with the DES-SV Cosmic Shear Cosmology paper~\citep{DES15}. 
The constraints from only using this configuration of Planck data when 
assuming a $w$CDM model are shown as the red contours in Fig.~\ref{fig:wCDM}.

With the inclusion of the DES $\gamt$ and $\wtheta$ measurements, we
were able to improve on the constraints on $\sigma_8$ and $w$ from
just Planck alone, which prefers $w\approx -1.5$ and $\sigma_8 \approx
1$. This is in part because DES provides modest constraints on $H_0$
%(about 15\% error) 
which help break the degeneracy between $h$
and $\Omega_m$  in the CMB. In addition, the Planck
dataset prefers higher values of $\sigma_8$ and $h$ than the DES data, 
such that in combination, the two probes carve out a
smaller area in parameter space. This produces strong constraints on
$w$ when the two datasets are combined. In combination with Planck, 
we find that $\Omega_m = 0.32\pm  0.02$, $\sigma_8 = 0.88 \pm  0.03$ 
and $w = -1.15 \pm  0.09$. 

\begin{figure*}
\centering
\begin{minipage}{.325\textwidth}
  \centering
  \includegraphics[width=1.0\linewidth]{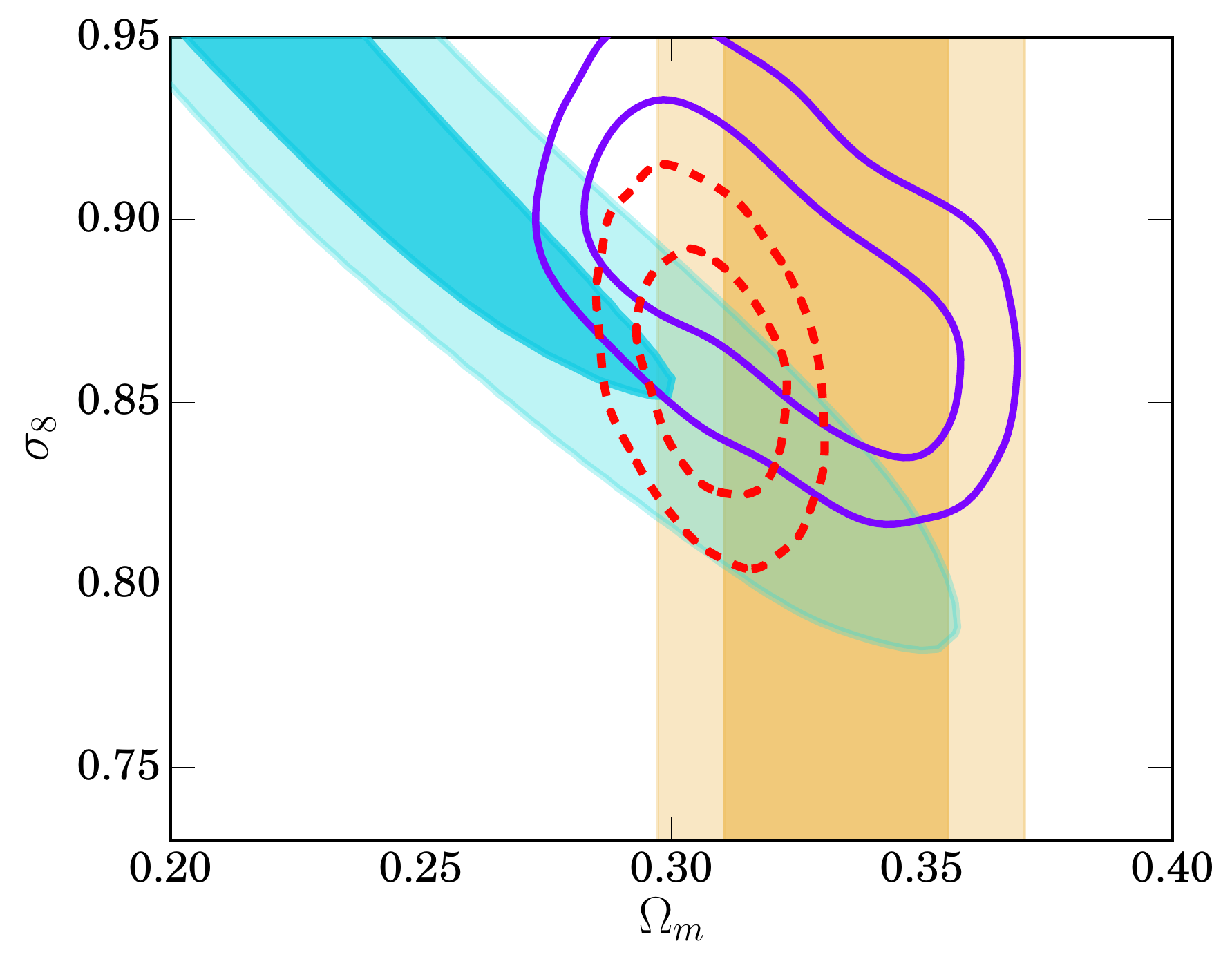} 
\end{minipage}%
\hfill
\begin{minipage}{.325\textwidth}
  \centering
  \includegraphics[width=1.0\linewidth]{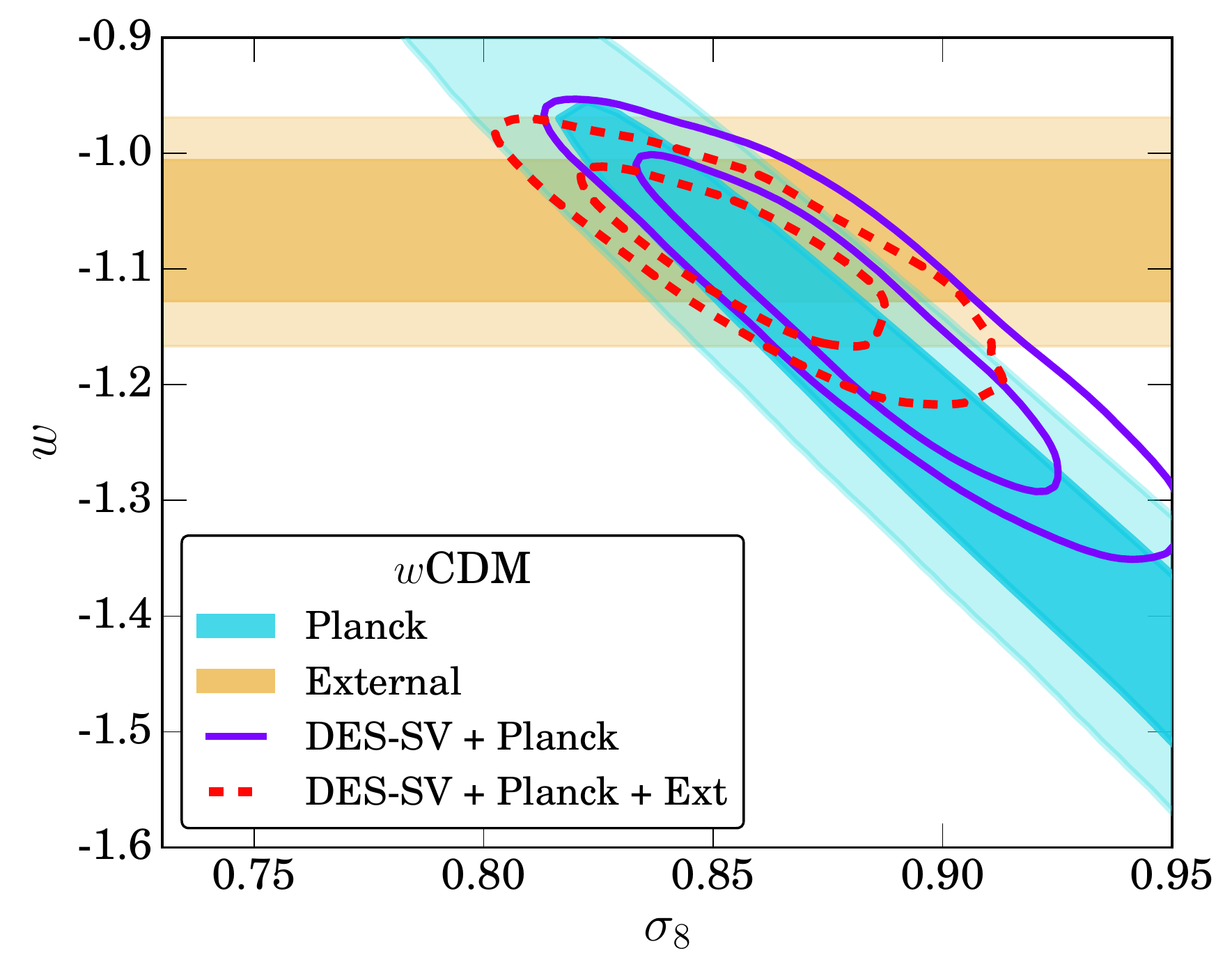} 
\end{minipage}
\hfill
\begin{minipage}{.325\textwidth}
  \centering
  \includegraphics[width=1.0\linewidth]{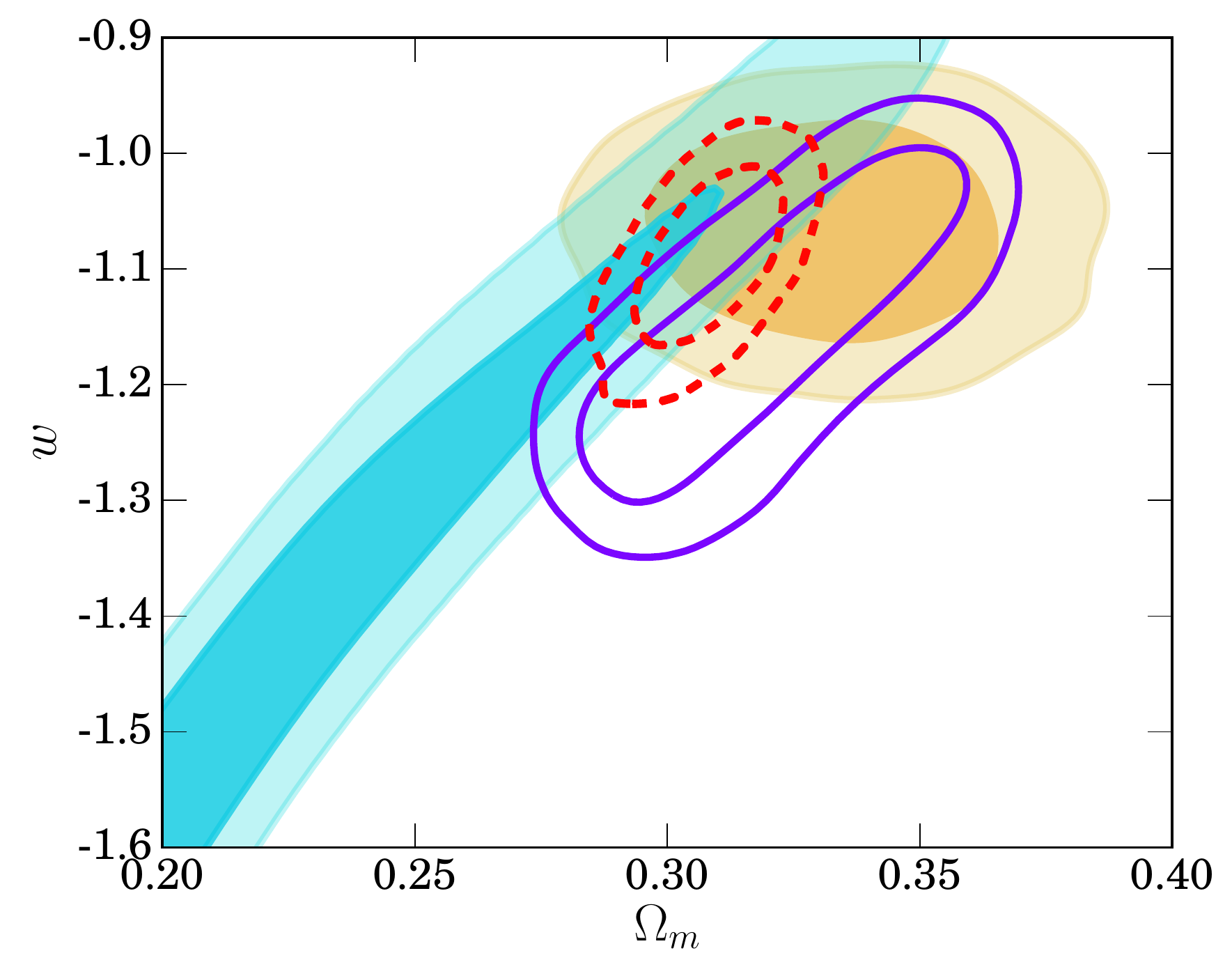} 
\end{minipage}
 \caption{Constraints on $\sigma_8$, $w$ and $\Omega_m$ using DES ($\wtheta$ x $\gamt$)
   in combination with Planck (solid purple) and DES in combination with Planck plus BAO, 
   SN Ia and $H_0$ measurements (dashed red). Also shown are the constraints from Planck only (filled blue) and BAO, 
   SN Ia and $H_0$ measurements only (filled yellow). }
 \label{fig:des_planck_ext_data}
\end{figure*}

Fig.~\ref{fig:des_planck_ext_data} shows the result of combining our 
measurements with additional data sets beyond the CMB. The other
probes that we consider are BAO measurements from 6dF~\citep{beutler11}, 
BOSS~\citep{anderson14, ross15}, Supernova type Ia measurements~\citep{betoule14}
and direct measurements of $H_0$~\citep{efstathiou14}. These data sets alone
give constraints of $\Omega_m = 0.33 \pm 0.02$ and $w = -1.07 \pm 0.06$ and no
constraint on $\sigma_8$ (the posterior distribution on $\sigma_8$ is fully informed by
the prior). 
Combining these data sets with DES and the CMB gives an improvement in precision 
and strengthens our results to $\Omega_m = 0.31 \pm  0.01$ and $\sigma_8 = 0.86 \pm 0.02$ 
and $w = -1.09 \pm 0.05$. 

%This is a substantial improvement over 
%constraints from using these external data sets alone, as shown in Fig.~\ref{fig:des_planck_ext_data}.

%%%%%%%%%%%%%%%
%\begin{comment}
%\end{comment}
%%%%%%%%%%%%%%%%%%%%%%%%%%%%%%%%%%%%%%%%%%%%%%

\subsection{Comparison with DES cosmic shear}
\citet{DES15} measured the 2-point shear correlations, for the same
DES-SV area and source catalogs. The best fitting cosmological
parameters in that work were $\sigma_8 = 0.81^{+0.16}_{-0.26}$ and $\Omega_m
=0.36^{+0.09}_{-0.21}$.  Figs. \ref{fig:comparison_shear_2pt} and \ref{fig:wCDM} show
the constraints from the analysis presented in this work on those parameters together with constraints from the shear 
2-point correlations for the $\Lambda$CDM and $w$CDM models, respectively. There
is very good agreement between the two analyses 
and a similar degeneracy direction in the $\Omega_m$ -- $\sigma_8$
plane as well. 

The shape of the contours for the two methods in Fig.~\ref{fig:comparison_shear_2pt} 
is somewhat different, with the cosmic shear contours being more elongated. 
We find that the slope $\alpha$ in the derived parameter S$_8 \equiv \sigma_8(\Omega_m/0.3)^\alpha$
is 0.16 for $\wtheta$ and $\gamt$ instead of 0.478 for cosmic shear.
In part because the covariance between $\Omega_m$ and $\sigma_8$ is weaker, the constraints on 
each parameter are slightly stronger for the $\wtheta$ and $\gamt$ case. The results in this analysis are
less sensitive to errors in the lensing shear and redshift distribution of source galaxies since these
do not impact $\wtheta$ at all, and additive errors in the shear cancel out of $\gamt$ at lowest order. 
On the  other hand, cosmic shear measurements are unaffected by errors in the galaxy biasing model 
and systematic errors in the measurement of galaxy clustering. Furthermore, the derived parameter S$_8$
is better constrained by DES cosmic shear. 
% S$_8 = 0.811^{+0.059}_{-0.060}$ (fiducial) for $\alpha = 0.478$
%than DES $\wtheta \times \gamt$: S$_8 = 0.736\pm{0.117}$. 
While there is significant complementarity in the two measurements, they are also
correlated because of the shared source galaxies. The combination of all three 
2-point functions taking into account  covariances is an important next step in the cosmological 
analysis of DES.
% our analysis with cosmic shear would gains us less constraining 
%power than expected, we obtain stronger measurements of $\sigma_8$ 
%and $\Omega_m$. While using cosmic shear with the same redshift binning
%in the shear catalog could provide a better handle on the photo-z and 
%shear calibration parameters, these are not dominant sources of uncertainty
%in our measurement. 

\subsection{Comparison with the literature}

A number of previous papers have considered the combination of
$\wtheta$ and $\gamt$ as probes of cosmology.  \citet{mandelbaum13}
perform an analysis with SDSS DR7 using luminous red galaxies as the
lenses and derive comparable constraints.  With some cosmological
parameters fixed, \citet{mandelbaum13} used a combination of three
lensing and angular clustering measurements in the redshift range 0$<
z < $0.5 to obtain $\sigma_8 = 0.76 \pm 0.08$ and $\Omega_m =
0.27^{+0.04}_{-0.03}$. 
%\citet{mandelbaum13} also uses compensated statistics that
%remove the non-linear contribution to $\wtheta$ and $\Delta\Sigma$.
%However, their analysis assumes that the covariance between the two
%probes and different redshift samples which share the same source
%catalog are small. %JC: This last assumption is more justified in their case, since source overlaps are much smaller
Several details of our analysis differ from \citet{mandelbaum13}, but the broad approach of employing a 
quasilinear analysis on large scales is similar and the results are consistent. 
%Furthermore, some cosmological parameters are
%held fixed in their analysis when using SDSS measurements alone that
%we have allowed to be free in
%Fig~\ref{fig:contours}. \citet{mandelbaum13} also finds a similar
%degeneracy is broken when observations of galaxy-galaxy lensing and
%$\wtheta$ are combined with the CMB.

\citet{cacciato13} also measure the tangential shear and angular
clustering from SDSS DR7 data, but differ in that they include small
scale clustering and consider a subset of the galaxy samples used
by~\citet{mandelbaum13}. They adopt a halo model approach which allows
them to extend their analysis to much smaller scales
than~\citet{mandelbaum13}, at the expense of requiring additional free
parameters and model ingredients that are calibrated with
simulations. With this small scale approach, \citet{cacciato13} obtain
$\Omega_m =0.278^{+0.023}_{-0.026}$ and $\sigma_8 =
0.763^{+0.064}_{-0.049}$, again consistent with our derived constraints.
%%JK: should we point out that this is in tension with mandelbaum to 
%% almost 2.5 sigma? 

Similarly, \citet{more15} use a halo model approach to calculate the
joint likelihood using galaxy clustering, galaxy-galaxy lensing and
galaxy abundance for the CMASS sample observed in BOSS using CFHTLenS
sources. They report that $\Omega_m = 0.31 \pm 0.02$ and
$\sigma_8 = 0.79 \pm 0.04$. Applying an HOD model motivates the
inclusion of small scale information in their cosmology fits.
%This and the additional information available in the
%galaxy abundance means that their constraints are tighter than ours by
%about a factor of 2. Like~\citet{mandelbaum13}, \citet{more15} also
%assumes CMB priors on the remaining cosmological parameters, while we
%took a more conservative approach and allowed them to vary in our
%analysis. 
In terms of number density and typical halo mass, the CMASS galaxies
used by \citet{more15} are closer to our \redmagic{}
sample than the LRGs in \citet{mandelbaum13}, but they all derive consistent cosmological constraints.

\section{Conclusions}
\label{sec:conclusion}

In this paper we have presented  cosmological constraints from the combination of large-scale structure and weak gravitational lensing in the Dark Energy Survey. Using a contiguous patch of 139 sq.~deg.~from the Science Verification period of observations, we have placed constraints on the matter density and the amplitude of fluctuations in the Universe as $\Omega_m = 0.31 \pm 0.09$ and $\sigma_8 = 0.74 \pm 0.13$, respectively. We also present joint constraints with CMB measurements from Planck, and additional low-redshift datasets. When allowing for a dark energy equation of state parameter $w$ different to the $\Lambda$CDM value of $-1$, we find DES data improve the constraints on $\sigma_8$ as well as $w$. 
%When using the full sample of \redmagic{} galaxies, we find that our constraints tighten from S$_8 = 0.74 \pm 0.12$ to S$_8 =  0.78 \pm 0.09$, for a fixed $\alpha = 0.16$. This implies that any improvements to the lower redshift catalog that would 
%allow for it to be included in the fiducial analysis would greatly enhance the constraining power of these probes. 
We leave a full tomographic analysis with multiple lens bins and a 
joint analysis with cosmic shear for future DES releases. 
%When we include constraints from the CMB, we measure $\Omega_m = 0.345 \pm 0.013$ and $\sigma_8 = 0.881 \pm 0.018$.

We have assessed the robustness of our results with respect to several variations in the choice of data vector, modelling and treatment of systematics. In particular, the results are stable under the use of two different shear catalogs, four different photo-$z$ codes and two different  estimators of the lensing signal. They also show consistency with the fiducial results when using a different lens bin, a different selection of angular scales or when adding a nonlinear galaxy bias parameter. 

The DES-SV region comprises only $\sim$3\% of the eventual survey
coverage, and we expect to greatly improve on our constraining power
with future data releases. For now, the analysis presented in this
paper is complementary to and provides a useful consistency check with the analysis of the shear
2-point function presented in~\citet{DES15}. These analyses validate the robust modelling of systematic errors and
galaxy bias, as well as the exhaustive testing of the shear
pipeline, photo-$z$ estimation and the \redmagic{} galaxy sample selection in the Dark Energy Survey.

\section*{Acknowledgments}

We are grateful for the extraordinary contributions of our CTIO colleagues and the DECam 
Construction, Commissioning and Science Verification teams in achieving the excellent 
instrument and telescope conditions that have made this work possible. The success of this 
project also relies critically on the expertise and dedication of the DES Data Management group.

\begin{comment}
CC and AA are supported by the Swiss National Science Foundation grants 
200021-149442 and 200021-143906. SB and JZ acknowledge support from a 
European Research Council Starting Grant with number 240672. 
DG was supported by
SFB-Transregio 33 `The Dark Universe' by the Deutsche
Forschungsgemeinschaft (DFG) and the DFG cluster of excellence `Origin
and Structure of the Universe'. 
FS acknowledges financial support provided by CAPES under contract No. 3171-13-2. 
OL acknowledges support from a European Research Council Advanced Grant FP7/291329
\end{comment}

Funding for the DES Projects has been provided by the U.S. Department of Energy, the U.S. National Science 
Foundation, the Ministry of Science and Education of Spain, the Science and Technology Facilities Council of 
the United Kingdom, the Higher Education Funding Council for England, the National Center for Supercomputing 
Applications at the University of Illinois at Urbana-Champaign, the Kavli Institute of Cosmological Physics 
at the University of Chicago, the Center for Cosmology and Astro-Particle Physics at the Ohio State University,
the Center for Particle Cosmology and the Warren Center at the University of Pennsylvania, 
the Mitchell Institute for Fundamental Physics and Astronomy at Texas A\&M University, Financiadora de 
Estudos e Projetos, Funda{\c c}{\~a}o Carlos Chagas Filho de Amparo {\`a} Pesquisa do Estado do Rio de 
Janeiro, Conselho Nacional de Desenvolvimento Cient{\'i}fico e Tecnol{\'o}gico and the Minist{\'e}rio da 
Ci{\^e}ncia e Tecnologia, the Deutsche Forschungsgemeinschaft and the Collaborating Institutions in the 
Dark Energy Survey. 

The DES data management system is supported by the National Science Foundation under Grant Number 
AST-1138766. The DES participants from Spanish institutions are partially supported by MINECO under 
grants AYA2012-39559, ESP2013-48274, FPA2013-47986, and Centro de Excelencia Severo Ochoa 
SEV-2012-0234, some of which include ERDF funds from the European Union.

The Collaborating Institutions are Argonne National Laboratory, the University of California at Santa Cruz, 
the University of Cambridge, Centro de Investigaciones Energeticas, Medioambientales y Tecnologicas-Madrid, 
the University of Chicago, University College London, the DES-Brazil Consortium, the Eidgen{\"o}ssische 
Technische Hochschule (ETH) Z{\"u}rich, Fermi National Accelerator Laboratory,
the University of Edinburgh, 
the University of Illinois at Urbana-Champaign, the Institut de Ci\`encies de l'Espai (IEEC/CSIC), 
the Institut de F\'{\i}sica d'Altes Energies, Lawrence Berkeley National Laboratory, the Ludwig-Maximilians 
Universit{\"a}t and the associated Excellence Cluster Universe, the University of Michigan, the National Optical 
Astronomy Observatory, the University of Nottingham, The Ohio State University, the University of Pennsylvania, 
the University of Portsmouth, SLAC National Accelerator Laboratory, Stanford University, the University of 
Sussex, and Texas A\&M University.

%%%%%%%%%%%%%%%

\appendix
\section{Excess Surface Density $\Delta\Sigma$}\label{sec:deltasigma}

\begin{figure}
\centering
\resizebox{80mm}{!}{\includegraphics{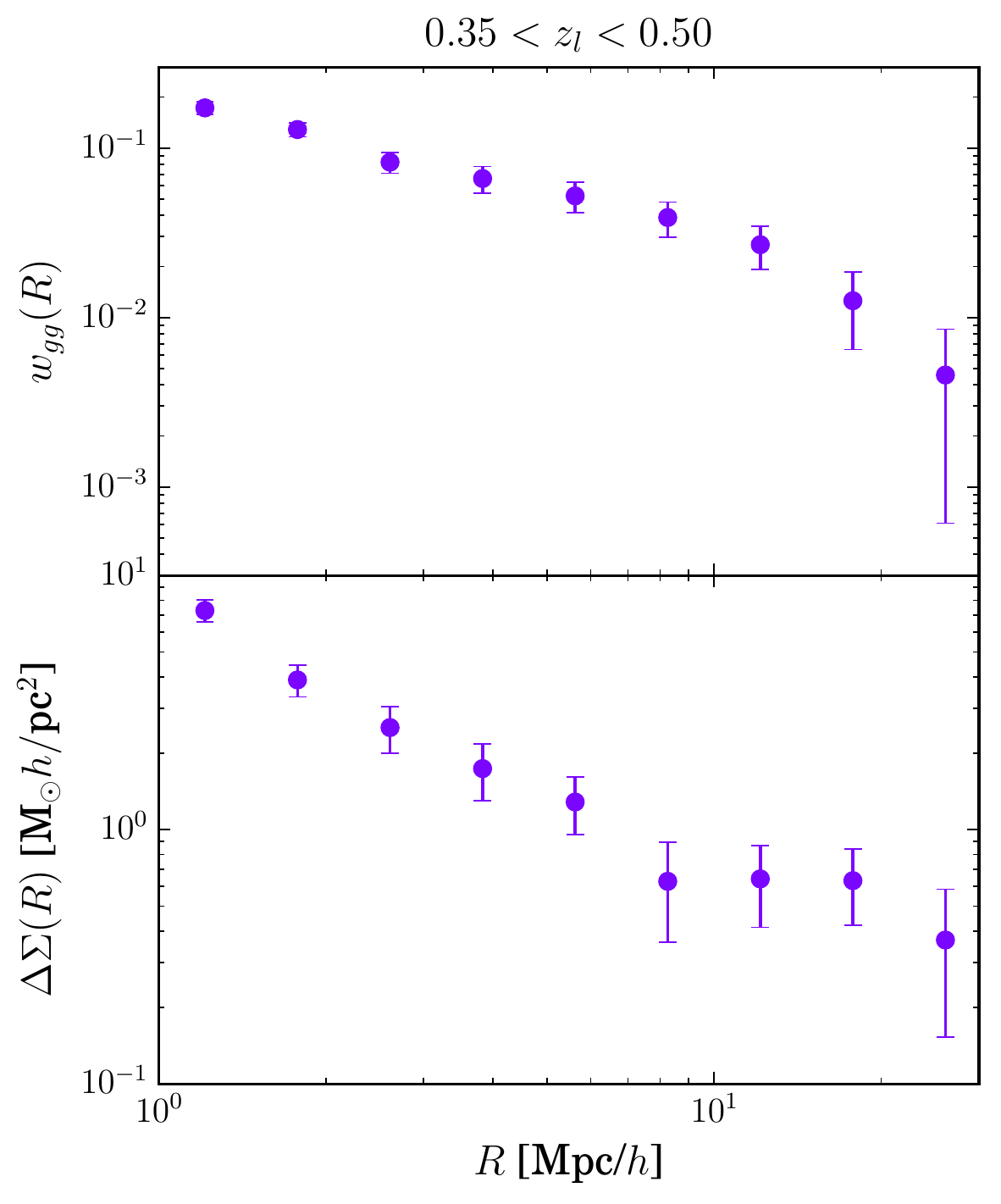}}
\caption{Same as our fiducial measurement plot in Fig.~\ref{fig:measurements}, but using
the alternative lensing estimator $\Delta\Sigma$.
In addition, the data are binned with respect to projected physical distance (R [Mpc$/h$]) rather than angle ($\theta$ [arcmins]).
The measurements are very similar to our fiducial results, as are the corresponding cosmological constraints in Fig.~\ref{fig:constraints_table}.}
\label{fig:deltasigma}
\end{figure}

In this section, we present complementary cosmology results obtained
for the fiducial \redmagic{} lens bin ($0.35<z<0.50$) by using the
excess surface density, $\Delta\Sigma(R)$ as a proxy for the
galaxy-galaxy lensing signal of \redmagic{} galaxies.
For this purpose, we define another lensing estimator that optimally weights each
lens-source pair of galaxies depending on the line-of-sight distance
separating them. This effectively downweights pairs of galaxies which
are very close and for which we expect a small lensing efficiency.
The observable is estimated from the measured shapes of background
galaxies as
\begin{equation} \label{eq:dsig_lens}
 \Delta\Sigma^{\rm lens}(R;\zl)=\frac{\sum_j 
\left[
\omega'_j \gamma_{t,j}(R) / \Sigma_{{\rm crit}, j}^{-1}(\zl, \zs)
\right]
}{\sum_j \omega'_j}
\end{equation}
where the summation $\sum_j$ goes over all the source galaxies in the
radial bin $R$, around all the lens galaxy positions, and the weight
factor for the $j$-th galaxy is given by
\be \label{eq:weight}
\omega'_j = \omega_j \, \Sigma_{{\rm crit}, j}^{-2}(\zl, \zs) \, .
\ee
Note that, in contrast with $\gamt$, for $\Delta\Sigma$ we bin
source galaxies according to radial distance $R$ in the region around
each lens galaxy, instead of angular scale $\theta$. In order to
estimate distances, we assume a flat $\Lambda$CDM model with
$\Omega_m=0.3$.  The weighting factor $\Sigma_{\rm crit}(\zl,\zs)$ is
computed as a function of lens and source redshifts for the assumed
cosmology as
\be \label{eq:sigma_crit}
\Sigma_{\rm crit} (\zl, \zs) = \frac{c^2}{4\pi G} \frac{D_A(\zs)}{D_A(\zl) D_A(\zl,\zs)} \, ,
\ee
where $\Sigma_{\rm crit}^{-1}(\zl,\zs)=0$ for $\zs<\zl$ and $D_A$ is
the angular diameter distance. We have checked that changes in the
assumed cosmology have little impact in the estimation of
$\Delta\Sigma$ so that they are not relevant for the analysis
presented in this work (see also \citealt{mandelbaum13}). Finally, just
as we do with tangential shear measurements, our final estimator
involves subtracting the contribution around random points, to which
now we assign redshifts randomly drawn from the real lens redshift
distribution.

Figure~\ref{fig:deltasigma} shows the clustering and the galaxy-galaxy
lensing signals, the latter using the alternative $\Delta\Sigma$
estimator, both binned according to projected radial distance $R$
around lenses.  In this case, we use all source galaxies available in
the {\ngmix} fiducial shear catalog and we weight each lens-source galaxy
pair according to their individual photometric redshifts so that
nearby pairs for which we expect a small lensing efficiency are
effectively downweighted. For the angular clustering, essentially the
same dataset is used in Fig.~\ref{fig:deltasigma} as for our fiducial
results pictured in Fig.~\ref{fig:measurements}. Thus, the two plots
are very similar, with the main difference being the range of scales
shown on the x-axis.

Our cosmological constraints obtained from fitting for
$\Delta\Sigma(R)$ and $w(R)$ are shown in
Fig.~\ref{fig:constraints_table}. These are consistent with our fiducial results, and show
tighter constraints on parameters like $\Omega_m$, due to the optimal
lens weighting and the larger number of source galaxies effectively
used. However, we do not use this estimator as the fiducial given the
larger uncertainties in the photo-$z$ modelling, which enters here on
a galaxy-by-galaxy basis unlike the tangential shear which only uses
the full stacked distribution.  Instead, we take the conservative
approach of using the tangential shear lensing measurements for the
fiducial case, which has better control over the potential photo-$z$
systematic effects.
%prevents us from using $\Delta\Sigma$ as a fiducial lensing estimator and stick to the most conservative tangential shear case. \cs{Premilinary! Rephrase that?}  

%%%%%%%%%%%%%%%%%%
\section{\texttt{ngmix} vs. \texttt{im3shape}}

\begin{figure}
\centering
\resizebox{90mm}{!}{\includegraphics{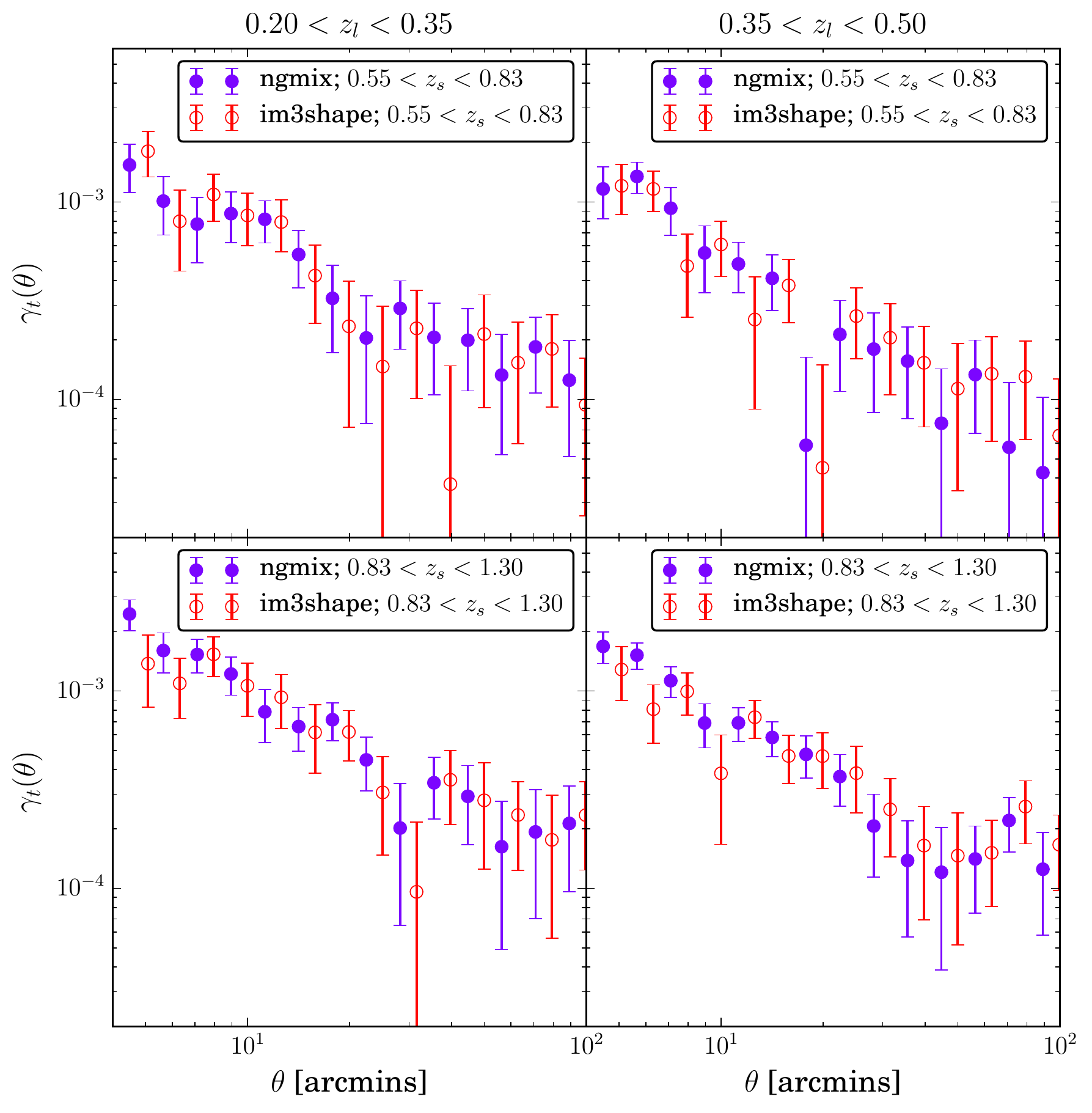}}
\caption{Comparison of the tangential shear signal using \texttt{ngmix} (solid purple
  circles) and \texttt{im3shape} (open red circles) shear pipelines.
  The result is shown for the two lens redshift bins (left and right columns) and
  the two source redshift bins (upper and lower rows) used in this work.
  For all bin combinations, the agreement between pipelines is excellent.}
  %JC: Modified caption
\label{fig:ngmix_im3shape}
\end{figure}

In Section~\ref{sec:shear_sys} we studied the consistency of the obtained cosmological constraints when using the two shear pipelines presented in
\citet{jarvis15}. In Fig.~\ref{fig:ngmix_im3shape} we show the actual comparison of the lensing measurements from the two shear pipelines, for all the different lens - source bin configurations.
The \texttt{im3shape} results are an excellent match to our fiducial measurements
with \texttt{ngmix} (shown earlier in Fig.~\ref{fig:measurements}).

\bibliographystyle{mn2e}
%\bibliographystyle{mn2e_eprint}

%\section*{Appendix}
%\renewcommand{\thesubsection}{\Alph{subsection}}

%%%%%%%%%%%%%%%%%%%
% Bibliography
%%%%%%%%%%%%%%%%%%%

%%%%%%%%%%%%%%%%%%%

\section*{Affiliations}
$^1$ Department of Physics and Astronomy, University of Pennsylvania, Philadelphia, PA 19104, USA\\
$^2$ Institut de F\'{\i}sica d'Altes Energies (IFAE), The Barcelona Institute of Science and Technology, Campus UAB, 08193 Bellaterra (Barcelona) Spain\\
$^3$ Center for Cosmology and Astro-Particle Physics, The Ohio State University, Columbus, OH 43210, USA\\
$^4$ Institut de Ci\`encies de l'Espai, IEEC-CSIC, Campus UAB, Carrer de Can Magrans, s/n,  08193 Bellaterra, Barcelona, Spain\\
$^5$ Jodrell Bank Center for Astrophysics, School of Physics and Astronomy, University of Manchester, Oxford Road, Manchester, M13 9PL, UK\\
$^6$ Department of Physics, ETH Zurich, Wolfgang-Pauli-Strasse 16, CH-8093 Zurich, Switzerland\\
$^7$ Department of Physics, Stanford University, 382 Via Pueblo Mall, Stanford, CA 94305, USA\\
$^8$ Kavli Institute for Particle Astrophysics \& Cosmology, P. O. Box 2450, Stanford University, Stanford, CA 94305, USA\\
$^9$ Fermi National Accelerator Laboratory, P. O. Box 500, Batavia, IL 60510, USA\\
$^{10}$ Kavli Institute for Cosmological Physics, University of Chicago, Chicago, IL 60637, USA\\
$^{11}$ Department of Physics, University of Chicago, 5640 South Ellis Avenue, Chicago, IL, 60637, USA\\
$^{12}$ Jet Propulsion Laboratory, California Institute of Technology, 4800 Oak Grove Dr., Pasadena, CA 91109, USA\\
$^{13}$ Institute of Astronomy, University of Cambridge, Madingley Road, Cambridge CB3 0HA, UK\\
$^{14}$ Kavli Institute for Cosmology, University of Cambridge, Madingley Road, Cambridge CB3 0HA, UK\\
$^{15}$ SLAC National Accelerator Laboratory, Menlo Park, CA 94025, USA\\
$^{16}$ Department of Physics, ETH Zurich, Wolfgang-Pauli-Strasse 16, CH-8093 Zurich, Switzerland\\
$^{17}$ Department of Physics \& Astronomy, University College London, Gower Street, London, WC1E 6BT, UK\\
$^{18}$ Instituci\'o Catalana de Recerca i Estudis Avan\c{c}ats, E-08010 Barcelona, Spain\\
$^{19}$ Department of Physics, University of Arizona, Tucson, AZ 85721, USA\\
$^{20}$ Brookhaven National Laboratory, Bldg 510, Upton, NY 11973, USA\\
$^{21}$ Cerro Tololo Inter-American Observatory, National Optical Astronomy Observatory, Casilla 603, La Serena, Chile\\
$^{22}$ Department of Physics and Electronics, Rhodes University, PO Box 94, Grahamstown, 6140, South Africa\\
$^{23}$ CNRS, UMR 7095, Institut d'Astrophysique de Paris, F-75014, Paris, France\\
$^{24}$ Sorbonne Universit\'es, UPMC Univ Paris 06, UMR 7095, Institut d'Astrophysique de Paris, F-75014, Paris, France\\
$^{25}$ Laborat\'orio Interinstitucional de e-Astronomia - LIneA, Rua Gal. Jos\'e Cristino 77, Rio de Janeiro, RJ - 20921-400, Brazil\\
$^{26}$ Observat\'orio Nacional, Rua Gal. Jos\'e Cristino 77, Rio de Janeiro, RJ - 20921-400, Brazil\\
$^{27}$ Department of Astronomy, University of Illinois, 1002 W. Green Street, Urbana, IL 61801, USA\\
$^{28}$ National Center for Supercomputing Applications, 1205 West Clark St., Urbana, IL 61801, USA\\
$^{29}$ Institute of Cosmology \& Gravitation, University of Portsmouth, Portsmouth, PO1 3FX, UK\\
$^{30}$ School of Physics and Astronomy, University of Southampton,  Southampton, SO17 1BJ, UK\\
$^{31}$ Excellence Cluster Universe, Boltzmannstr.\ 2, 85748 Garching, Germany\\
$^{32}$ Faculty of Physics, Ludwig-Maximilians-Universit{\"a}t, Scheinerstr. 1, 81679 Munich, Germany\\
$^{33}$ Department of Astronomy, University of Michigan, Ann Arbor, MI 48109, USA\\
$^{34}$ Department of Physics, University of Michigan, Ann Arbor, MI 48109, USA\\
$^{35}$ Department of Physics, The Ohio State University, Columbus, OH 43210, USA\\
$^{36}$ Australian Astronomical Observatory, North Ryde, NSW 2113, Australia\\
$^{37}$ Departamento de F\'{\i}sica Matem\'atica,  Instituto de F\'{\i}sica, Universidade de S\~ao Paulo,  CP 66318, CEP 05314-970, S\~ao Paulo, SP,  Brazil\\
$^{38}$ George P. and Cynthia Woods Mitchell Institute for Fundamental Physics and Astronomy, and Department of Physics and Astronomy, Texas A\&M University, College Station, TX 77843,  USA\\
$^{39}$ Department of Astronomy, The Ohio State University, Columbus, OH 43210, USA\\
$^{40}$ Department of Astrophysical Sciences, Princeton University, Peyton Hall, Princeton, NJ 08544, USA\\
$^{41}$ Max Planck Institute for Extraterrestrial Physics, Giessenbachstrasse, 85748 Garching, Germany
$^{42}$ Department of Physics and Astronomy, Pevensey Building, University of Sussex, Brighton, BN1 9QH, UK\\
$^{43}$ Centro de Investigaciones Energ\'eticas, Medioambientales y Tecnol\'ogicas (CIEMAT), Madrid, Spain\\
$^{44}$ ICTP South American Institute for Fundamental Research\\ Instituto de F\'{\i}sica Te\'orica, Universidade Estadual Paulista, S\~ao Paulo, Brazil\\
$^{45}$ Argonne National Laboratory, 9700 South Cass Avenue, Lemont, IL 60439, USA\\

\bsp
\label{lastpage}

\end{document}